\documentclass{article}

\usepackage{iclr2027_conference,times}
\usepackage{enumitem}
\usepackage[utf8]{inputenc}
\usepackage[T1]{fontenc}
\usepackage[hidelinks]{hyperref}
\usepackage{url}
\usepackage{booktabs}
\usepackage{amsmath,amssymb,amsfonts}
\usepackage{graphicx}
\usepackage{subcaption}
\usepackage{wrapfig}
\usepackage{float}
\usepackage{placeins}
\usepackage{algorithm}
\usepackage{algpseudocode}
\usepackage{xcolor}
\usepackage{microtype}
\usepackage{wrapfig}
\newcommand{\Cbb}{\mathbb{C}}
\newcommand{\Ebb}{\mathbb{E}}
\newcommand{\bs}[1]{\boldsymbol{#1}}
\newcommand{\xb}{\bs{x}}
\newcommand{\yb}{\bs{y}}
\newcommand{\hb}{\bs{h}}
\newcommand{\db}{\bs{d}}
\newcommand{\method}{G-Wave}

\title{G-Wave: Bit-to-Waveform Generation for\\Adaptive Wireless Communications}

\author{\textbf{Hanwen Zhang\textsuperscript{1}\quad
Yunfan Leng\textsuperscript{3}\quad
Zheng Shi\textsuperscript{1}}\\
\textbf{Bo Yuan\textsuperscript{2}\quad
Zhenlin An\textsuperscript{3}\quad
Haijian Sun\textsuperscript{1}}\\[4pt]
\normalfont\small\textsuperscript{1}School of Electrical and Computer Engineering, University of Georgia\\
\normalfont\small\textsuperscript{2}Department of Electrical and Computer Engineering, Rutgers University\\
\normalfont\small\textsuperscript{3}School of Computing, University of Georgia}
\hypersetup{
pdftitle={G-Wave: Bit-to-Waveform Generation for Adaptive Wireless Communications},
pdfauthor={Hanwen Zhang, Yunfan Leng, Zheng Shi, Bo Yuan, Zhenlin An, Haijian Sun}}

\iclrfinalcopy

\begin{document}

\maketitle
\lhead{Preprint}

\begin{abstract}
Wireless waveforms are traditionally hand-designed for general propagation conditions, rather than adapted to each specific environment. Existing neural transceivers seek such adaptation, but often rely on relatively simple architectures with limited long-range modeling capability. More expressive Transformer-based generative architectures are difficult to apply directly because wireless communication demands extremely accurate bit recovery, often with BER below $0.1\%$, while providing no ground-truth waveform supervision. To address these limitations, we introduce \method{}, a channel-conditioned generative transceiver that directly synthesizes environment-adaptive physical time-domain waveforms from bits. \method{} consists of Transformer-based transmitter and receiver models equipped with a multiscale architecture that combines long-range sequence modeling with fine-grained signal synthesis to meet the stringent precision required for communication. The two models are jointly optimized end-to-end using only bit-reconstruction error, allowing the waveform itself to be discovered without waveform labels. Across diverse 3GPP wireless environments, \method{} achieves lower BER than the SOTA neural transceiver (DeepOFDM) and a strong conventional  baseline (OFDM) under  estimated channel information, reducing mean BER by $42.4\%$ and $72.6\%$, respectively.

\end{abstract}

\section{Introduction}
\label{sec:intro}

\begin{wrapfigure}{r}{0.55\textwidth}
    \vspace{-8pt}
    \centering
    \includegraphics[width=\linewidth]{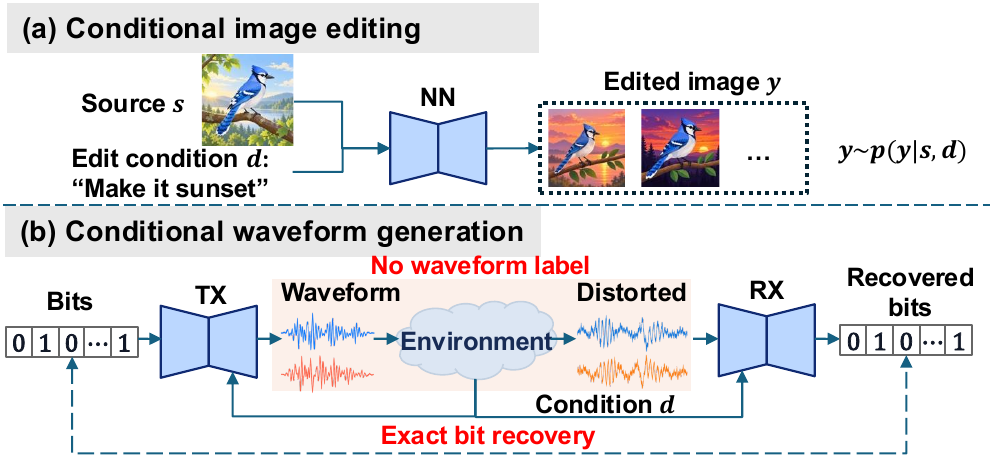}
    \captionsetup{
        font=footnotesize,
        labelfont=bf,
        skip=2pt
    }
    \caption{
        Comparison between conventional conditional generation and
        bit-to-waveform-to-bit wireless communication.
    }
    \label{fig:problem-comparison}
    \vspace{-4mm}
\end{wrapfigure}
Wireless communication is a cornerstone of modern society, enabling information exchange among vast numbers of interconnected devices~\citep{itu2025facts}. At the heart of this connectivity is \emph{waveform design}, which determines how digital information is embedded in physical signals for transmission over wireless channels~\citep{tse2005fundamentals}. This representation directly affects how reliably the information can be recovered after the transmitted signals undergo channel distortion and noise~\citep{aoudia2022waveform}. Modern waveforms such as orthogonal frequency-division multiplexing (OFDM) are the result of decades of theory development and engineering refinement, combining modulation, orthogonal frequency-domain signaling, and Fourier transforms into a robust communication framework~\citep{weinstein1971data,tse2005fundamentals}. However, these waveforms are largely designed as general-purpose solutions across diverse propagation conditions, rather than being optimized for each specific wireless environment. This motivates a question: \textit{can we train a model to automatically discover effective waveforms tailored to a given environment and communication objective?}

Recent advances in generative modeling offer a promising path toward such environment-adaptive waveform design through condition-guided generation and structural discovery~\citep{rombach2022highresolution,peebles2023scalable,watson2023denovo}. From our perspective, environment-adaptive waveform design can be viewed as a condition-guided sequence transformation problem, where the bit sequence specifies the information to be conveyed, while the wireless environment controls how that information is physically realized as a time-domain waveform. Despite this conceptual similarity, existing conditional generative models for images, text, and audio cannot be directly transferred to communication waveform synthesis, which imposes substantially stricter requirements (Fig.~\ref{fig:problem-comparison}). (1) Every payload bit must be recovered accurately (i.e., error rate < 0.1\%), whereas conventional generative tasks are typically evaluated at the semantic or perceptual level and can often tolerate stochastic variations in the generated output. (2) The communication chain traverses heterogeneous  domains, from binary coded bits to complex-valued continuous waveforms and finally back to bits, whose statistical structures and functional roles differ substantially. (3) No ground-truth waveform specifies the desired physical realization for a given payload and propagation environment; the transmitter and receiver must therefore jointly discover the waveform representation using only the bit-reconstruction error backpropagated through the communication chain. 
\begin{figure}
\centering
\includegraphics[width=\linewidth]{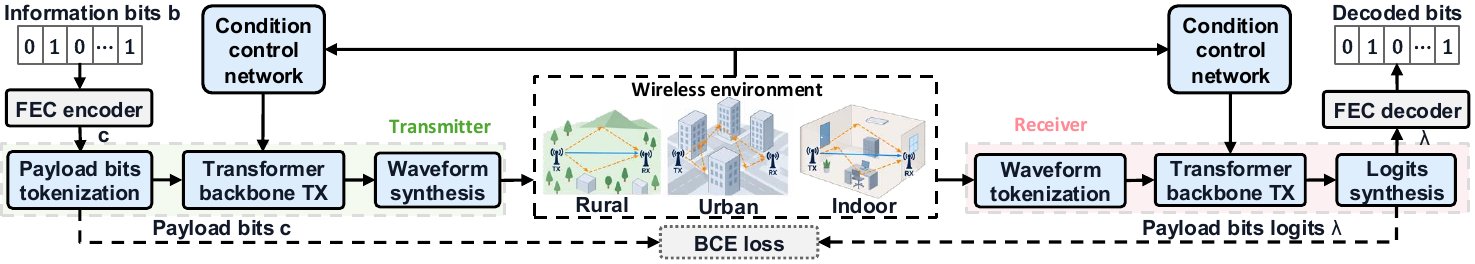}\vspace{-3mm}
\caption{\textbf{Overview of \method{}.} Environmental conditions modulate waveform synthesis and recovery; the illustrated propagation scenes denote alternative operating conditions. The transmitter and receiver are trained jointly through the bit-reconstruction loss.}
\label{fig:arch}
\vspace{-6mm}
\end{figure}
We address these challenges with \method{}, a wireless-environment-conditioned end-to-end generative transceiver (Fig.~\ref{fig:arch}). The key contributions are:

\begin{itemize}[leftmargin=*]
\item \textbf{End-to-end waveform-label-free bit-to-waveform generation.} We develop a fully differentiable generative transceiver that directly maps coded bits to physical waveforms and jointly optimizes the transmitter and receiver through the wireless channel. This end-to-end formulation removes the requirement for waveform labels, allowing the signaling waveform to be discovered only from bit-reconstruction supervision. Moreover, by generating waveforms directly from bits without predefined modulation formats, \method{} exposes a broader waveform design space for discovering better environment-adaptive signaling structures and decreasing the bit error rate.

\item \textbf{Multiscale latent representation and precise synthesis.} We develop a residual multiscale interface that combines global context and fine temporal features within a conditioned Transformer~\citep{tian2024visual,feng2026msdformer}. This coarse-to-fine representation enables channel information to guide latent feature processing at multiple temporal resolutions. Through domain-specific projection and synthesis heads, the resulting features support channel-adaptive IQ waveform synthesis from coded bits and estimation of coded-bit logits from received IQ samples effectively.

\item \textbf{Strong empirical performance with interpretability and ablations.}
Under a controlled single-link evaluation, \method{} achieves lower mean information-bit error rate (BER) than DeepOFDM~\citep{ankireddy2025deepofdm}, DeepOFW~\citep{greidi2026deepofw}, and conventional OFDM under both perfect and estimated channel state information (CSI). On the mixed-channel test distribution with estimated CSI, it reduces mean information-bit BER across the evaluated $E_b/N_0$ points by $42.4\%$ relative to DeepOFDM and $72.6\%$ relative to conventional OFDM. Across the mixed-channel and scenario-specific evaluations, \method{} maintains reliable bit recovery under diverse propagation conditions, including estimated CSI, while synthesizing waveforms that respond to the supplied channel condition.
\end{itemize}
\vspace{-4mm}
\section{Related work}
\label{sec:related}
\textbf{End-to-End and Environment-Conditioned Learned Communication.}
Early learned communication systems modeled the transmitter and receiver as an autoencoder and demonstrated end-to-end optimization mainly over relatively simple channel models~\citep{oshea2017introduction,dorner2018deep,aoudia2019modelfree,cammerer2020trainable}. For more challenging frequency-selective channels, subsequent work commonly introduced neural networks into established communication pipelines, replacing or optimizing selected modules~\citep{aoudia2022waveform,felix2018ofdm,aoudia2021end,honkala2021deeprx,ankireddy2025deepofdm,wu2022channeladaptive}. Feedback-based approaches instead use receiver-side observations or channel-related feedback to adapt subsequent transmitted representations~\citep{kim2018deepcode,song2020quantizedfeedback,shao2023attentioncode,kurka2020deepjsccf,wu2024transformerfeedback}. Another line of work uses hypernetworks to map  CSI into the parameters of neural transceivers~\citep{goutay2020deep,zhang2022adaptive,greidi2026deepofw}. Nevertheless, these approaches generally retain designer-specified physical-layer signal structures and rely on relatively simple mechanisms for incorporating channel conditions, limiting their flexibility in channel-adaptive waveform generation. In contrast, \method{} learns the bit-to-waveform mapping directly in a latent feature space, where wireless environment conditions are injected to modulate the learned representation and guide the synthesis of channel-adaptive complex time-domain waveforms.

\textbf{Generative Models for Wireless Communication.}
Generative models have been applied to wireless communication mainly for RF signal and data synthesis~\citep{shi2019spoofing,karunaratne2021outlier,chi2024rfdiffusion,kaasaragadda2025reformer}, wireless channel modeling and estimation~\citep{balevi2021wideband,arvinte2023scorebased,fesl2024diffusion,sengupta2023channel,zhou2025diffusion}, and semantic-level communication~\citep{qiao2024latency,erdemir2023generative,wang2025diffcom,fu2026land,chen2026bits}. These approaches primarily synthesize RF data for augmentation and downstream tasks, infer channel-related information, or reconstruct semantically equivalent content, rather than directly generating a physical waveform that encodes a designated bit sequence. In contrast, our goal is precise bit-to-waveform generation, which lies beyond the scope of these existing approaches.

\section{Preliminaries: math model for wireless communication}
\label{sec:preliminaries}
A typical wireless communication system consists of a transmitter and a receiver connected through a wireless propagation channel. At the transmitter, bits information is first encoded into a complex-valued time-domain waveform represented by its in-phase and quadrature (IQ) samples. These digital IQ samples are converted by the RF front end into a real-valued waveform and transmitted over the air. During propagation, environmental effects, including multipath, attenuation, and noise, distort the transmitted waveform and produce a received IQ sequence. The receiver processes this sequence to characterize the propagation effects and recover the transmitted bits. This communication process can be described mathematically through the following sequence of mappings. Before waveform generation, a forward error correction (FEC) encoder maps the information bits $\bs{b}\in\{0,1\}^{K}$ to a coded sequence with structured redundancy,
\begin{equation}
\bs{c}=f_{\mathrm{enc}}(\bs{b})\in\{0,1\}^{C},
\label{eq:link-coding}
\end{equation}
where $K$ and $C$ denote the information-bit and coded-bit lengths, respectively, with coding rate $K/C<1$. The redundancy introduced by FEC enables the receiver to correct transmission errors during decoding. Then, the transmitter maps $\bs{c}$ to an unnormalized time-domain complex-baseband waveform $\widetilde{\xb}$ and rescales it to satisfy the average transmit-power constraint:
\begin{equation}
\widetilde{\xb}=f_{\eta}^{\mathrm{TX}}(\bs{c})\in\Cbb^N,
\qquad
\xb=\rho\widetilde{\xb}.
\label{eq:link-tx}
\end{equation}
where $f_{\eta}^{\mathrm{TX}}$ denotes the transmitter waveform mapping parameterized by $\eta$, and $N$ is the length of the transmitted waveform sequence $\xb$. The normalization factor is $\rho=\sqrt{NP/\|\widetilde{\xb}\|_2^2}$, which ensures that the transmitted waveform has average transmit power $P$. During over-the-air propagation, multipath propagation and additive complex Gaussian noise $\bs{n}$ distort the transmitted waveform, yielding the received baseband sequence
\begin{equation}
\yb=\hb\circledast\xb+\bs{n},
\qquad
\bs{n}\sim\mathcal{CN}(0,N_0\bs{I}),
\label{eq:link-channel}
\end{equation}
where $\hb\in\Cbb^{L_h}$ denotes the complex channel impulse response (CIR), $\circledast$ denotes linear convolution, and $N_0$ is the noise variance. The receiver processes the received sequence $\yb$ and produces a soft estimate for each coded bit:
\begin{equation}
\bs{\lambda}=f_{\xi}^{\mathrm{RX}}(\yb)\in\mathbb{R}^{C},
\qquad
\widehat{\bs{b}}=f_{\mathrm{dec}}(\bs{\lambda}),
\label{eq:link-rx}
\end{equation}
where $f_{\xi}^{\mathrm{RX}}$ denotes the receiver processing pipeline parameterized by $\xi$, encompassing receiver-side operations such as channel estimation, equalization, and soft bit inference, and $\bs{\lambda}$ contains the resulting coded-bit logits. The function $f_{\mathrm{dec}}$ denotes the FEC decoder paired with the encoder $f_{\mathrm{enc}}$, which uses these soft estimates to recover the information bits $\widehat{\bs{b}}$. The reliability of the complete communication pipeline is measured by the information-bit error rate (BER), defined as $\mathrm{BER}=d_{\mathrm H}(\bs{b},\widehat{\bs{b}})/K$, where $d_{\mathrm H}(\cdot,\cdot)$ denotes the Hamming distance.

\section{\method: Integer-Aligned Conditional Generative Transceiver}
\label{sec:method}

\subsection{System model and end-to-end pipeline}
\label{sec:system}

\method{} parameterizes the transmitter and receiver as environment-conditioned neural networks:
\begin{equation}
\widetilde{\xb}
=
f_{\theta}^{\mathrm{NN\text{-}TX}}(\bs{c},\db),
\qquad
\bs{\lambda}
=
f_{\phi}^{\mathrm{NN\text{-}RX}}(\yb,\db),
\label{eq:neural-transceiver}
\end{equation}
where $\db$ denotes the wireless environment descriptor constructed from the channel state information (CSI) available to the transceiver and the noise-variance information. $\theta$ and $\phi$ are the learnable parameters of the neural transmitter and receiver, respectively, and are jointly optimized end-to-end. Figure~\ref{fig:arch} provides an overview of the \method{} architecture. The neural transmitter and receiver are independently parameterized but share a similar high-level design, consisting of three main components: a residual multiscale representation and synthesis interface, a bidirectional Transformer, and an environment-conditioning mechanism.

\subsection{Residual multiscale representation and synthesis interface}
\label{sec:tokenizer}

To enable precise bit-preserving waveform synthesis, we design a fully differentiable multiscale interface that can be optimized jointly with the transceiver. Waveform generation requires coordinating information over the full frame while resolving local sample variations; bit recovery similarly combines broad received context with fine signal features. We therefore represent latent sequences at several temporal resolutions, process their tokens jointly, and fuse their outputs on a common grid. The multicarrier organization of OFDM and coarse-to-fine generative modeling~\citep{tian2024visual,feng2026msdformer} motivate this design, whose scales represent temporal resolutions in the learned latent space. Gradient sensitivity, token swapping, and branch-removal analyses in Appendix~\ref{app:scale-contrib} characterize their contributions to waveform synthesis and bit recovery.

\begin{figure}[!t]
\centering
\includegraphics[width=1.0\linewidth]{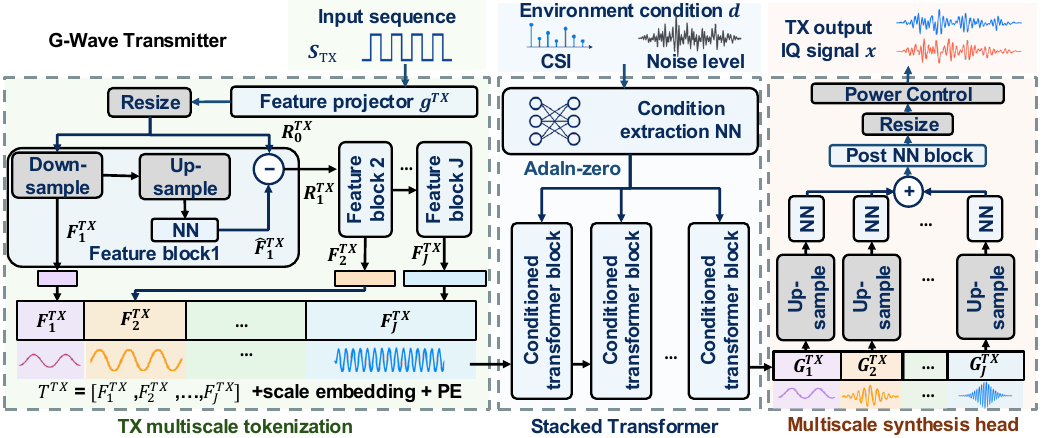}\vspace{-3mm}
\caption{\textbf{Multiscale neural transmitter of \method{}.} Residual multiscale representation extracts coarse-to-fine features, which are concatenated, augmented with scale and positional embeddings, and processed by an environment-conditioned bidirectional Transformer. The multiscale synthesis head then reconstructs and fuses the transformed features across scales before mapping them to the time-domain waveform.}\vspace{-4mm}
\label{fig:multiscale-interface}
\end{figure}

\textbf{Transmitter multiscale representation.}
To facilitate subsequent neural processing, the transmitter first converts the binary coded-bit sequence into a zero-centered numerical representation. Specifically, we center the coded bits as $\bs{s}_{\mathrm{TX}}=\bs{c}-\tfrac{1}{2}\bs{1}_{C}$, yielding a one-channel sequence in $\{-\tfrac{1}{2},\tfrac{1}{2}\}^{C\times 1}$. We next project the centered bit sequence into a latent feature space and align its sequence length with the transmitter latent grid used for waveform synthesis. This is implemented by a 1D convolutional feature projector followed by sequence-length resizing:
\begin{equation}
\bs{R}_0^{\mathrm{TX}} = \mathcal{I}_{C\rightarrow L^{\mathrm{TX}}} \left( g_{\mathrm{TX}}(\bs{s}_{\mathrm{TX}}) \right) \in\mathbb{R}^{L^{\mathrm{TX}}\times D},
\label{eq:gwave-tx-feature-projection}
\end{equation}
where $g_{\mathrm{TX}}$ lifts the scalar coded-bit sequence into a $D$-dimensional latent feature space, and $\mathcal{I}_{C\rightarrow L^{\mathrm{TX}}}$ resizes the sequence length from $C$ to $L^{\mathrm{TX}}$. The resulting length-$L^{\mathrm{TX}}$ latent grid is used for the subsequent multiscale representation and is chosen to support integer-aligned waveform synthesis. Starting from $\bs{R}_0^{\mathrm{TX}}$, we decompose the latent representation into a coarse-to-fine hierarchy of $J$ scales with lengths $\ell_1^{\mathrm{TX}}<\cdots<\ell_J^{\mathrm{TX}}=L^{\mathrm{TX}}$, where $\ell_1^{\mathrm{TX}}$ denotes the coarsest scale and $\ell_J^{\mathrm{TX}}$ corresponds to the full latent resolution. At each scale $j=1,\ldots,J-1$, the current residual $\bs{R}_{j-1}^{\mathrm{TX}}\in\mathbb{R}^{L^{\mathrm{TX}}\times D}$ is first resized to length $\ell_j^{\mathrm{TX}}$ to obtain the scale-specific feature $\bs{F}_j^{\mathrm{TX}}\in\mathbb{R}^{\ell_j^{\mathrm{TX}}\times D}$. The extracted feature is then projected back to the full latent resolution to form a reconstructed feature $\widehat{\bs{F}}_j^{\mathrm{TX}}\in\mathbb{R}^{L^{\mathrm{TX}}\times D}$, which is subtracted from the current residual:
\begin{equation}
\begin{aligned}
\bs{F}_j^{\mathrm{TX}}&=\mathcal{I}_{L^{\mathrm{TX}}\rightarrow \ell_j^{\mathrm{TX}}}(\bs{R}_{j-1}^{\mathrm{TX}}),\quad
\widehat{\bs{F}}_j^{\mathrm{TX}}=\phi_j^{\mathrm{TX}}\!\left(\mathcal{I}_{\ell_j^{\mathrm{TX}}\rightarrow L^{\mathrm{TX}}}(\bs{F}_j^{\mathrm{TX}})\right),\\
\bs{R}_j^{\mathrm{TX}}&=\bs{R}_{j-1}^{\mathrm{TX}}-\widehat{\bs{F}}_j^{\mathrm{TX}},\quad j=1,\ldots,J-1.
\end{aligned}
\label{eq:gwave-multiscale}
\end{equation}

After residual extraction, the multiscale features are concatenated from coarse to fine to form \(Q^{\mathrm{TX}}=\sum_{j=1}^{J}\ell_j^{\mathrm{TX}}\) tokens. We add a learned scale embedding shared by tokens from the same scale, together with positional encoding (PE), before feeding the sequence into the bidirectional Transformer. The scale embedding identifies each token's temporal resolution, while PE provides its positional information. The resulting multiscale tokens are then jointly processed by the bidirectional Transformer, with the wireless environment condition injected through AdaLN-Zero~\citep{peebles2023scalable} to coordinate information across temporal positions and scales. As illustrated in Figure~\ref{fig:multiscale-interface}, this stage comprises the transmitter-side multiscale representation and environment-conditioned feature transformation, producing coordinated multiscale tokens for subsequent waveform synthesis.

\textbf{Transmitter waveform-synthesis head.} To reconstruct the physical waveform from the Transformer output, we follow the same multiresolution principle used in the input representation. The coordinated features provide complementary resolutions for representing global and local waveform structure. Rather than decoding these scales serially, we reconstruct them in parallel on a common latent grid and combine their contributions to synthesize the final waveform. Specifically, the Transformer output $\bs{G}^{\mathrm{TX}}\in\mathbb{R}^{Q^{\mathrm{TX}}\times D}$ is partitioned according to the original scale boundaries as $\bs{G}^{\mathrm{TX}}=[\bs{G}^{\mathrm{TX},(1)},\ldots,\bs{G}^{\mathrm{TX},(J)}]$, where $\bs{G}^{\mathrm{TX},(m)}\in\mathbb{R}^{\ell_m^{\mathrm{TX}}\times D}$ denotes the feature block associated with the $m$-th scale. Each block is resized to the full latent length $L^{\mathrm{TX}}$ and processed by a scale-specific branch $\psi_m^{\mathrm{TX}}$, after which the reconstructed features are summed and jointly refined by a post-fusion block:
\begin{equation}
\bs{U}^{\mathrm{TX}}=\psi_{\mathrm{post}}^{\mathrm{TX}}\!\left(\sum_{m=1}^{J}\psi_m^{\mathrm{TX}}\!\left(\mathcal{I}_{\ell_m^{\mathrm{TX}}\rightarrow L^{\mathrm{TX}}}(\bs{G}^{\mathrm{TX},(m)})\right)\right)\in\mathbb{R}^{L^{\mathrm{TX}}\times D}.
\label{eq:gwave-head}
\end{equation}
The fused representation $\bs{U}^{\mathrm{TX}}$ is finally mapped to the physical waveform through an integer-aligned readout. Let $p_{\mathrm{TX}}=N/L^{\mathrm{TX}}$ denote the integer output ratio, such that each latent token directly generates $p_{\mathrm{TX}}$ complex IQ samples. The readout produces the corresponding real and imaginary components, which are paired into complex samples and concatenated along the temporal dimension to form the unnormalized waveform $\widetilde{\xb}\in\Cbb^N$.

\begin{figure}[!t]
\centering
\includegraphics[width=1.0\linewidth]{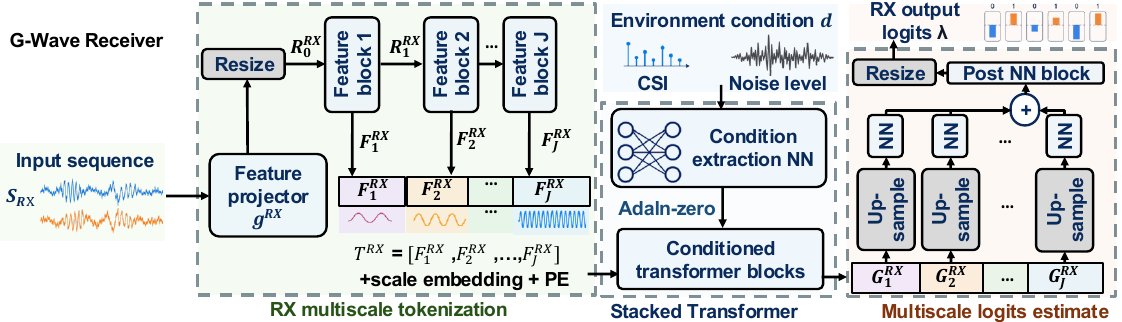}\vspace{-3mm}
\caption{\textbf{Receiver architecture of \method{}. }The received IQ sequence is projected into receiver-specific multiscale tokens, conditioned on CSI and noise variance through AdaLN-Zero Transformer blocks, and fused by the multiscale output head to produce coded-bit logits.}\vspace{-4mm}
\label{fig:multiscale-rx}
\end{figure}

\textbf{Receiver-side adaptation.} As illustrated in Figure~\ref{fig:multiscale-rx}, the receiver mirrors the high-level architecture of the transmitter while adapting its input and output interfaces to waveform-to-bit inference. The received complex waveform is first represented by stacking its real and imaginary components along the channel dimension, $\bs{s}_{\mathrm{RX}}=[\operatorname{Re}(\yb),\operatorname{Im}(\yb)]\in\mathbb{R}^{N_{\mathrm{RX}}\times 2}$, and projected into the receiver latent space. The resulting sequence is decomposed into coarse-to-fine multiscale features, concatenated into receiver tokens, and processed by a receiver-specific bidirectional Transformer, with the wireless environment condition injected through AdaLN-Zero in the same manner as at the transmitter. At the output, the transformed tokens are split by scale and lifted to a common latent resolution through the multiscale synthesis head; the scale-wise features are then fused, refined by a post-fusion neural block, and mapped to coded-bit logits through an integer-aligned readout. Let $p_{\mathrm{RX}}=C/L^{\mathrm{RX}}$, such that each latent position produces $p_{\mathrm{RX}}$ logits and the complete receiver output is $\bs{\lambda}\in\mathbb{R}^{C}$. These logits provide soft estimates of the transmitted coded bits and are subsequently used by the fixed channel decoder for information-bit recovery. Detailed pseudocode for the shared multiscale representation and output head is provided in Algorithms~\ref{alg:tokenizer} and~\ref{alg:head} in Appendix~\ref{app:multiscale-algorithms}.

\textbf{Environment conditioning.} To adapt the synthesized waveform and receiver inference to the current wireless environment, we condition both sides of the transceiver on CSI and noise-variance information. We represent CSI by the CIR taps $\hb$ and convert these quantities into a real-valued environment descriptor:
\begin{equation}
\db=\left[\nu(N_0),\operatorname{Re}(\hb)^{\mathsf T},\operatorname{Im}(\hb)^{\mathsf T}\right]^{\mathsf T}\in\mathbb{R}^{1+2L_h},
\label{eq:environment-descriptor}
\end{equation}
where $\nu(\cdot)$ denotes a normalized logarithmic transformation of the noise variance. Under estimated CSI, $\widehat{\hb}$ and $\widehat{N}_0$ replace $\hb$ and $N_0$ only in the descriptor; physical propagation still uses $\hb$. To effectively inject the environmental information into the latent representation, each side $a\in\{\mathrm{TX},\mathrm{RX}\}$ employs a small condition-extraction MLP $q^{a}$ to map the environment descriptor $\db$ to a condition embedding $\bs{e}^{a}=q^{a}(\db)$. Here $\bs{e}^{a}$ is the resulting condition embedding used to modulate the corresponding encoder-only bidirectional Transformer through AdaLN-Zero. PE provides positional information for the multiscale sequence, while bidirectional self-attention captures dependencies across temporal positions and scales. The transmitter and receiver use independent conditioning networks and Transformer parameters. Appendix~\ref{app:learning-objective} provides an information-theoretic rationale for our training objective and discusses the connection between learned waveform design and channel-capacity limits.

\textbf{End-to-end training and practical deployment.} Since no target waveform is available for either the transmitter output or the received signal, the complete neural transceiver is trained directly through a differentiable wireless channel. At each training iteration, we sample a coded-bit sequence, a propagation channel, a noise variance $N_0$, and a noise realization $\bs{n}$. The transmitter synthesizes a waveform without waveform-level supervision, which is propagated through the sampled channel and processed by the receiver to produce coded-bit logits $\bs{\lambda}$. The transmitter and receiver are then jointly optimized using only the coded-bit reconstruction loss $\mathcal{L}=\mathcal{L}_{\mathrm{BCE}}(\bs{c},\bs{\lambda})$. The channel coding scheme is fixed rather than learned: the encoder provides the coded sequence $\bs{c}$, while the channel decoder is used only for information-bit evaluation. Consequently, gradients propagate directly from the coded-bit loss through the receiver, differentiable channel, and synthesized waveform to the transmitter, without requiring intermediate waveform labels or separately pretrained components. At inference, the trained transmitter and receiver share an environment descriptor to synthesize the waveform and recover coded-bit logits; practical deployment details are given in Appendix~\ref{app:deployment}.

\section{Experiments}
\label{sec:experiments}

\subsection{Experimental setup}
\label{sec:setup}

\textbf{Link configuration.} We evaluate a single-user, single-antenna frequency-selective link. Each frame contains 384 information bits encoded by a rate-$1/2$ 5G LDPC code into 768 coded bits. The transmitter emits 490 complex time-domain samples. Linear convolution with a 19-tap channel produces 508 received samples, which are zero-padded to 512 samples before neural reception. The sampling rate is 480\,kHz and the carrier frequency is $3.5$\,GHz. Propagation follows the clustered delay line (CDL) model in 3GPP TR~38.901~\citep{3gpp38901}, with channel normalization disabled.

\textbf{Model and training.} The \method{} transmitter and receiver use multiscale latent lengths $(1,7,49,245)$ and $(4,16,64,256)$, respectively. Both sides contain a six-layer bidirectional Transformer with feature width 256, four attention heads, scale embeddings, PE, and AdaLN-Zero conditioning. We optimize it for 120,000 steps with Adam, batch size 16, learning rate $2\times10^{-4}$, and global gradient clipping at 1.0. At each training step, information bits are sampled i.i.d. from $\operatorname{Bernoulli}(1/2)$. For each batch, the CDL profile is drawn uniformly from A--E and its RMS delay spread is drawn log-uniformly from $0.1$ to $0.6\,\mu$s; terminal speed is fixed at $0$\,m/s. Training $E_b/N_0$ is sampled uniformly from $[0,15]$\,dB.

\textbf{Baselines.} We compare \method{} against three representative baselines: conventional OFDM, DeepOFDM~\citep{ankireddy2025deepofdm}, and DeepOFW~\citep{greidi2026deepofw}. Conventional OFDM uses QPSK modulation and LMMSE equalization with either perfect CSI or pilot-based LS channel estimates. DeepOFDM jointly learns a convolutional modulator and neural receiver while retaining a predefined OFDM resource grid; for channel-aware comparison, we concatenate CSI with its transmitter grid features. DeepOFW instead uses a CSI-conditioned parameter generator to produce the waveform matrix and detector coefficients of a structured transceiver. All neural baselines are retrained under the same channel distribution and evaluated with the same test protocol. Every method carries the same 384 information bits in 490 samples at the same sampling rate and frame-energy budget, and uses the same LDPC decoder. Explicit OFDM pilots and cyclic prefixes occupy part of this fixed frame budget; \method{} learns its signal representation within the same budget.

\textbf{Evaluation protocol.} Unless stated otherwise, methods are evaluated on paired channel and noise realizations, with at least 16,000 frames per $E_b/N_0$ point and additional samples for selected low-error points. BER denotes information-bit BER after 20 LDPC decoding iterations; coded-bit BER is measured from the receiver's hard decisions before LDPC decoding. Figure~\ref{fig:ber-csi}, Table~\ref{tab:cdl-mode-ber}, and their associated mean BERs use the nine-point grid $\{0,2,\ldots,14,15\}$\,dB. The supplementary coded/info BER comparison and block-error analysis use the full $0$--$15$\,dB sweep in $1$\,dB increments, as specified in Appendix~\ref{app:additional-comparisons}. For estimated CSI, learned methods share the same 19-tap CIR estimate $\widehat{\hb}$ from a common OFDM pilot observation. A mean BER averages the pointwise BERs with equal weights. Boldface marks the lowest BER in each table column, including ties; a dash in a BER entry denotes no observed post-decoding errors.

\subsection{End-to-end BER performance}
\label{sec:main}

\begin{figure*}[!htbp]
\centering
\includegraphics[width=0.84\textwidth]{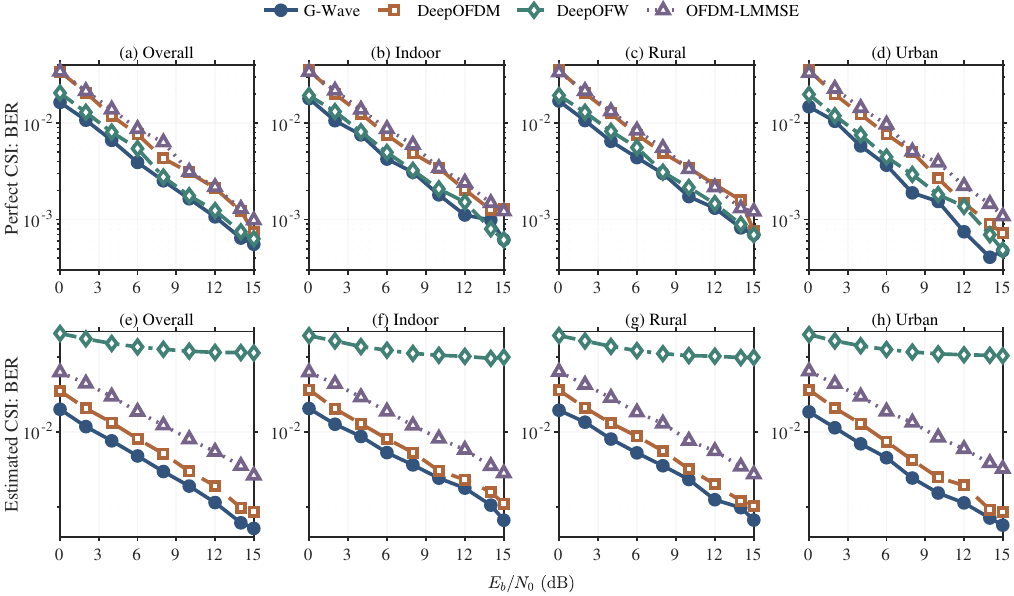}\vspace{-3mm}
\caption{End-to-end information-bit BER under perfect CSI (a)--(d) and estimated CSI (e)--(h). Columns show the overall CDL setting, indoor office, RMa, and UMa, respectively.} \vspace{-5mm}
\label{fig:ber-csi}
\end{figure*}

We first conduct an aggregate evaluation to evaluate the overall communication reliability of \method{} across diverse propagation conditions and determine whether its gains persist beyond any particular channel profile or delay-spread regime. \method{} and all baselines are evaluated over a mixed test distribution with randomly sampled CDL-A--E profiles and delay spreads across the considered $E_b/N_0$ range. We report results under both perfect and estimated CSI, with the latter examining whether the performance gains remain when channel-state information must be obtained through estimation. Figure~\ref{fig:ber-csi} summarizes the resulting BER performance.
\vspace{-3mm}

\method{} achieves the lowest mean information-bit BER on the mixed CDL-A--E test distribution under both CSI settings. Averaged over the evaluated $E_b/N_0$ points, its BER is $4.90\times10^{-3}$ with perfect CSI and $5.74\times10^{-3}$ with estimated CSI; the corresponding values for DeepOFDM are $9.52\times10^{-3}$ and $9.96\times10^{-3}$. \method{} thus retains low BER and its performance advantage with estimated CSI, indicating robustness to errors in the environmental descriptor. DeepOFW, by contrast, has mean BER $6.01\times10^{-3}$ with perfect CSI but $1.41\times10^{-1}$ with estimated CSI.

\textbf{Scenario-specific generalization.} We next evaluate three scenario-inspired delay-spread regimes from 3GPP TR~38.901~\citep{3gpp38901}: 20--60\,ns for indoor office, 32--153\,ns for rural macro (RMa), and 93--1148\,ns for urban macro (UMa). For each frame, we sample CDL-A--E uniformly and draw the RMS delay spread log-uniformly within the corresponding scenario range while keeping the trained transceivers fixed. As shown in Figure~\ref{fig:ber-csi}, \method{} achieves the lowest mean BER in each scenario under both CSI settings. Averaging over the evaluated $E_b/N_0$ points and weighting the three scenarios equally, its mean BER is $4.95\times10^{-3}$ under perfect CSI, compared with $5.90\times10^{-3}$ for DeepOFW. With estimated CSI, the corresponding mean is $5.94\times10^{-3}$ for \method{}, versus $1.01\times10^{-2}$ for DeepOFDM, $2.11\times10^{-2}$ for conventional OFDM, and $1.28\times10^{-1}$ for DeepOFW. Together, these results indicate that \method{} provides a more effective and robust transceiver design across the evaluated propagation conditions, particularly when only estimated CSI is available.

\textbf{Per-CDL-profile performance.} To evaluate the consistency of \method{} across a diverse set of standardized channel profiles, we separately test CDL-A--E and report the information-bit BER in Table~\ref{tab:cdl-mode-ber}. \method{} achieves the lowest mean BER on CDL-A--C under both perfect and estimated CSI. On CDL-D/E, several methods have no observed post-decoding bit errors over the evaluated frames. These results suggest that \method{} maintains reliable BER under each of the evaluated CDL profiles. Additional tests on unseen ray-traced channels, under user mobility, and before LDPC decoding further support the reliability of \method{}; detailed results are provided in Appendices~\ref{app:additional-comparisons} and~\ref{app:generalization}.

\begin{table}[H]
\centering
\caption{Mean information-bit BER over the evaluated $E_b/N_0$ points for each CDL profile.}
\label{tab:cdl-mode-ber}

\scriptsize
\setlength{\tabcolsep}{2.0pt}
\begin{tabular}{@{}l*{10}{c}@{}}
\toprule
& \multicolumn{10}{c}{Information-bit BER ($\times 10^{-3}$) $\downarrow$} \\
\cmidrule(lr){2-11}
Method & \multicolumn{2}{c}{\shortstack{CDL-A\\(Perfect/Estimated)}} & \multicolumn{2}{c}{\shortstack{CDL-B\\(Perfect/Estimated)}} & \multicolumn{2}{c}{\shortstack{CDL-C\\(Perfect/Estimated)}} & \multicolumn{2}{c}{\shortstack{CDL-D\\(Perfect/Estimated)}} & \multicolumn{2}{c}{\shortstack{CDL-E\\(Perfect/Estimated)}} \\
\midrule
\textbf{\method{} (Ours)} & \textbf{6.93} & \textbf{8.39} & \textbf{7.85} & \textbf{9.65} & \textbf{9.30} & \textbf{11.0} & \textbf{\textemdash} & \textbf{\textemdash} & \textbf{\textemdash} & \textbf{\textemdash} \\
DeepOFDM & 13.6 & 14.6 & 18.1 & 19.3 & 17.8 & 18.6 & $2.35{\times}10^{-4}$ & $1.18{\times}10^{-3}$ & \textbf{\textemdash} & \textbf{\textemdash} \\
DeepOFW & 8.35 & 123 & 10.6 & 125 & 11.2 & 124 & \textbf{\textemdash} & 132 & \textbf{\textemdash} & 141 \\
OFDM-LMMSE & 14.3 & 29.4 & 19.9 & 40.8 & 19.2 & 37.7 & $1.03{\times}10^{-3}$ & 0.246 & \textbf{\textemdash} & 0.210 \\
\bottomrule
\end{tabular}
\end{table}
\vspace{-5mm}

\subsection{Channel-conditioned waveform adaptation}
\label{sec:channel-adaptation} 
\begin{figure*}[!htbp]
\centering
\includegraphics[width=0.97\textwidth]{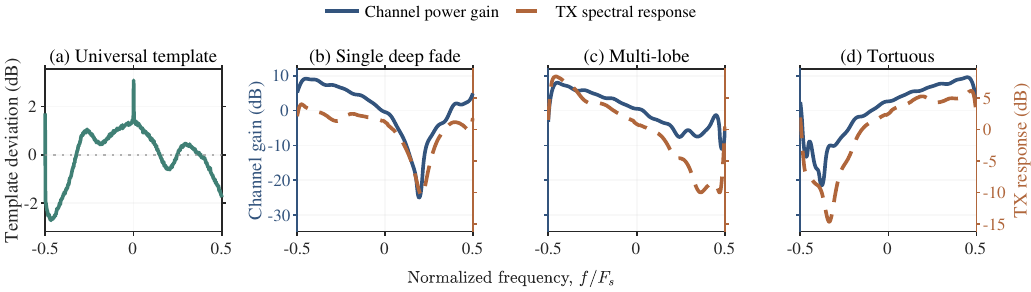}\vspace{-3mm}
\caption{Channel-conditioned spectral adaptation learned by \method{}. (a) Universal transmit-power template shared across channel realizations. Template-normalized TX spectral responses for three selected NLOS realizations: (b) CDL-B at $0.54\,\mu$s with a single deep fade; (c) CDL-A at $0.36\,\mu$s with a multi-lobe gain profile; and (d) CDL-C at $0.35\,\mu$s with a tortuous gain profile.}
\label{fig:wave-response}
\end{figure*}

\textbf{Shared spectral structure.} To evaluate channel-specific adaptation, we first characterize the spectral structure shared across generated waveforms. Let $X(f)$ denote the frequency-domain representation of the transmitted waveform $\xb$, where $f$ denotes normalized frequency. We estimate a universal transmit-power template as $T_{\mathrm{raw}}(f)=\mathbb{E}|X(f)|^2$ using 256 randomly sampled channels and 64 payloads per channel. As shown in Figure~\ref{fig:wave-response}(a), the waveforms exhibit a common power pattern, while the coherent shared component $|\mathbb{E}X(f)|^2$ accounts for only $0.074\%$ of the total power. This suggests that the generated waveforms are not dominated by a fixed preference for particular frequency regions; instead, most of the transmit-spectrum variation is shaped by the specific channel condition.

\textbf{Channel-specific spectral adaptation.} To better understand the BER gains of \method{} over conventional OFDM, we examine how the transmit spectrum changes with channel quality. Using the shared spectral template as a reference, for each test channel we compute the payload-averaged spectrum $P_{\hb}(f)=\mathbb{E}_{\bs{c}}|X(f)|^2$ and its relative response $R_{\mathrm{dB}}(f)=10\log_{10}\!\left(P_{\hb}(f)/T_{\mathrm{raw}}(f)\right)$. Figure~\ref{fig:wave-response}(b)--(d) compares this response with the channel gain for three representative channel realizations. The gain response correlations are $0.93$, $0.89$, and $0.93$ respectively. Across all cases, \method{} allocates more transmit power to frequency regions with stronger channel gain while suppressing deeply faded regions, thereby directing the limited transmit energy toward more favorable parts of the channel. Such adaptive power allocation helps reduce severe signal degradation in deeply faded frequencies and facilitates more reliable delivery to the receiver. Further spectral-response results across CDL profiles and delay spreads are provided in Appendix~\ref{app:txchan-grid}. Per-scale contributions and waveform-encoding/CSI-perturbation analyses are reported in Appendices~\ref{app:scale-contrib} and~\ref{app:influence}, respectively. Appendix~\ref{app:additional-comparisons} provides additional comparisons with OFDM using CSI-based water-filling. These results suggest that the performance gains of \method{} extend beyond power allocation to its learned bit-to-waveform representation. The observed spectral redistribution is one manifestation of this channel-adaptive waveform design.

\subsection{Ablation study}
\label{sec:ablation}

The full \method{} model combines the residual multiscale, AdaLN-Zero conditioning, PE, and direct time-domain waveform generation. Each ablation changes one design dimension relative to this reference. \emph{Single-scale} removes the serial coarse-to-fine residual decomposition and its associated tokenizer branches while retaining a single-scale neural interface. \emph{No scale embedding} deletes only the additive scale-label vectors from the TX and RX tokenizers. \emph{Concatenative conditioning} replaces layer-wise AdaLN-Zero modulation with broadcast condition features followed by dense reprojection, whereas \emph{No conditioning} zeros the environment descriptor at both endpoints. \emph{Frequency-domain I/O} preserves the neural backbone but inserts a unitary FFT/IFFT interface so that the network reads and writes frequency-domain coefficients. \emph{OFDM resource-grid I/O} further constrains waveform generation to a prescribed OFDM grid with IFFT and cyclic-prefix framing. All variants use the same channel distribution, frame-energy constraint, and paired perfect-CSI evaluation protocol. Table~\ref{tab:ablation} summarizes the results.

Replacing the multiscale representation with a single-scale variant or removing scale embeddings degrades BER. The learned coarse scales also affect waveform synthesis directly: their combined waveform-gradient share is $58.9\%$, and removing individual coarse TX branches increases BER (Appendix~\ref{app:scale-contrib}). Weaker or absent condition injection likewise reduces reliability, as do frequency-domain and OFDM-grid constraints. Together, these results support the integration of latent multiscale processing, layer-wise conditioning, and direct time-domain synthesis.

\begin{table}[!htbp]
\centering
\caption{Ablation under perfect CSI. Mean BER over the $0$--$15$\,dB sweep; $\times$ Full divides each variant's mean by the full model's mean.}
\label{tab:ablation}
\small
\setlength{\tabcolsep}{6pt}
\begin{tabular}{@{}lcccc@{}}
\toprule
& \multicolumn{2}{c}{Coded-bit BER $\downarrow$} & \multicolumn{2}{c}{Information-bit BER $\downarrow$} \\
\cmidrule(lr){2-3}\cmidrule(lr){4-5}
Variant & Mean & $\times$ Full & Mean & $\times$ Full \\
\midrule
\textbf{\method{}}
    & $\boldsymbol{7.04{\times}10^{-3}}$ & 1.00 & $\boldsymbol{4.57{\times}10^{-3}}$ & 1.00 \\
Single-scale
    & $7.79{\times}10^{-3}$ & 1.11 & $5.07{\times}10^{-3}$ & 1.11 \\
No scale embedding
    & $8.10{\times}10^{-3}$ & 1.15 & $5.21{\times}10^{-3}$ & 1.14 \\
Concatenative conditioning
    & $8.09{\times}10^{-3}$ & 1.15 & $5.24{\times}10^{-3}$ & 1.15 \\
No conditioning
    & $1.26{\times}10^{-2}$ & 1.79 & $9.22{\times}10^{-3}$ & 2.02 \\
Frequency-domain I/O
    & $7.78{\times}10^{-3}$ & 1.11 & $5.04{\times}10^{-3}$ & 1.10 \\
OFDM resource-grid I/O
    & $2.06{\times}10^{-2}$ & 2.93 & $9.69{\times}10^{-3}$ & 2.12 \\
\bottomrule
\end{tabular}
\end{table}

\subsection{Computational cost}
\label{sec:complexity}
We compare the computational cost of \method{} with the retrained DeepOFDM and DeepOFW baselines under the same frame settings. \method{} contains $17.59$\,M parameters and requires $11.0$\,GFLOP per frame; the corresponding values are $26.67$\,M and $23.9$\,GFLOP for DeepOFDM, and $17.68$\,M and $0.07$\,GFLOP for DeepOFW. On an NVIDIA A100 at batch size one, TX+RX inference takes $3.68$\,ms for \method{} and $3.23$\,ms for DeepOFDM with XLA compilation, while DeepOFW takes $2.00$\,ms using its optimized GRU implementation. Relative to DeepOFDM, \method{} achieves improved BER with fewer parameters and operations and similar optimized latency. Appendix~\ref{app:complexity} reports all baselines, execution settings, and memory measurements.

\section{Conclusion}
\label{sec:conclusion}

We introduced \method{}, a new end-to-end generative transceiver that adapts waveform generation to diverse wireless channel conditions and enables highly reliable bit transmission with very low BER. Its key technical contributions are a Transformer-based multiscale bit-to-waveform architecture for precise long-range signal modeling and an end-to-end bit-supervised training framework that jointly learns waveform generation and recovery without waveform labels. Beyond wireless bit transmission, we envision that this framework could inspire generative learning systems for other applications that require both adaptive signal synthesis and extremely high output accuracy, such as secure communications, cybersecurity, and other precision-critical learning tasks.
\clearpage
\subsection*{AI use statement}
Generative AI tools assisted with manuscript drafting and editing, literature search, feedback on the methodology and experimental design, preparation of illustrative figure drafts and plotting code, and discussion of experimental results. The authors reviewed and revised AI-assisted materials and take responsibility for the final text, code, figures, analyses, and claims.

\subsection*{Ethics statement}
This work evaluates learned waveform design in simulated, single-user wireless links using standardized and site-specific channel models and randomly generated bits. It does not involve human participants or personal data.

\subsection*{Reproducibility statement}
Section~\ref{sec:setup} specifies the link configuration, channel sampling, model and training settings, baselines, and evaluation protocol. Appendix~\ref{app:multiscale-algorithms} gives pseudocode for the multiscale components. Appendices~\ref{app:additional-comparisons}--\ref{app:complexity} document the supplementary baselines, error metrics, ray-tracing and mobility evaluations, and computational profiling. We use randomly generated bits, simulated CDL channels, and the cited public ray-tracing resources. We will publicly release the implementation and evaluation code after the paper is accepted.

\bibliography{references}
\bibliographystyle{iclr2027_conference}

\appendix
\section{Multiscale tokenizer and synthesis algorithms}
\label{app:multiscale-algorithms}

Algorithms~\ref{alg:tokenizer} and~\ref{alg:head} give the shared TX/RX steps for the multiscale tokenizer and output head, respectively. The side index $a$ selects the corresponding input, latent lengths, and learned parameters.

\noindent\begin{minipage}[t][2.30in][t]{0.485\linewidth}
\hrule height 0.8pt
\setlength{\abovecaptionskip}{4pt}
\setlength{\belowcaptionskip}{3pt}
\captionsetup{hypcap=false}
\captionof{algorithm}{Shared multiscale tokenizer}
\label{alg:tokenizer}
\hrule height 0.4pt
\vspace{0.35em}
\footnotesize
\begin{algorithmic}[1]
\Require side $a\in\{\mathrm{TX},\mathrm{RX}\}$; input $\bs{s}_{a}$ of length $L_{\mathrm{in}}^{a}$
\Require latent length $L^{a}$; scales $\ell_1^{a}<\cdots<\ell_J^{a}=L^{a}$
\State $\bs{R}_0^{a}\gets\mathcal{I}_{L_{\mathrm{in}}^{a}\rightarrow L^{a}}(g^{a}(\bs{s}_{a}))$
\For{$j=1,\ldots,J-1$}
  \State $\bs{F}_j^{a}\gets\mathcal{I}_{L^{a}\rightarrow\ell_j^{a}}(\bs{R}_{j-1}^{a})$
  \State $\widehat{\bs{F}}_j^{a}\gets\phi_j^{a}(\mathcal{I}_{\ell_j^{a}\rightarrow L^{a}}(\bs{F}_j^{a}))$
  \State $\bs{R}_j^{a}\gets\bs{R}_{j-1}^{a}-\widehat{\bs{F}}_j^{a}$
\EndFor
\State $\bs{F}_J^{a}\gets\bs{R}_{J-1}^{a}$
\State $\bs{T}^{a}\gets[\bs{F}_1^{a}\Vert\cdots\Vert\bs{F}_J^{a}]+\bs{E}_{\mathrm{scale}}^{a}+\bs{E}_{\mathrm{pos}}^{a}$
\State \Return $\bs{T}^{a}$ and $\{\ell_j^{a}\}_{j=1}^{J}$
\end{algorithmic}
\vfill
\hrule height 0.8pt
\end{minipage}\hfill
\begin{minipage}[t][2.30in][t]{0.485\linewidth}
\hrule height 0.8pt
\setlength{\abovecaptionskip}{4pt}
\setlength{\belowcaptionskip}{3pt}
\captionsetup{hypcap=false}
\captionof{algorithm}{Shared multiscale output head}
\label{alg:head}
\hrule height 0.4pt
\vspace{0.35em}
\footnotesize
\begin{algorithmic}[1]
\Require side $a\in\{\mathrm{TX},\mathrm{RX}\}$; coordinated tokens $\bs{G}^{a}$
\Require scales $\{\ell_m^{a}\}_{m=1}^{J}$; latent length $L^{a}$
\Require $p_{\mathrm{TX}}=N/L^{\mathrm{TX}}$; $p_{\mathrm{RX}}=C/L^{\mathrm{RX}}$
\State $\{\bs{G}^{a,(m)}\}_{m=1}^{J}\gets\Call{SplitByScale}{\bs{G}^{a},\{\ell_m^{a}\}}$
\State $\bs{U}^{a}\gets\psi_{\mathrm{post}}^{a}\!\left(\sum_{m=1}^{J}\psi_m^{a}\!\left(\mathcal{I}_{\ell_m^{a}\rightarrow L^{a}}(\bs{G}^{a,(m)})\right)\right)$
\If{$a=\mathrm{TX}$}
  \State $\bs{o}^{a}\gets\operatorname{ComplexReshape}(\operatorname{Linear}_{2p_{\mathrm{TX}}}(\bs{U}^{a}))$
\Else
  \State $\bs{o}^{a}\gets\operatorname{Reshape}(\operatorname{Linear}_{p_{\mathrm{RX}}}(\bs{U}^{a}))$
\EndIf
\State \Return $\bs{o}^{a}$, where $\bs{o}^{\mathrm{TX}}=\widetilde{\xb}$ and $\bs{o}^{\mathrm{RX}}=\bs{\lambda}$
\end{algorithmic}
\vfill
\hrule height 0.8pt
\end{minipage}

\section{Inference and practical deployment}
\label{app:deployment}

This appendix details the inference and deployment procedure for \method{}. During deployment, the neural transmitter and receiver use fixed trained parameters, while their condition embeddings incorporate the current environmental information.

\paragraph{Condition acquisition.} At the beginning of a condition-update interval, the receiver estimates the channel impulse response $\widehat{\hb}$ and noise variance $\widehat{N}_0$. A conventional estimator can obtain these quantities from a known preamble or OFDM pilot symbols. The receiver then constructs $\widehat{\db}=[\nu(\widehat{N}_0),\operatorname{Re}(\widehat{\hb})^{\mathsf T},\operatorname{Im}(\widehat{\hb})^{\mathsf T}]^{\mathsf T}$ using the same ordering and normalization as during training, and \method{} uses this fixed-length descriptor as its condition. Perfect-CSI evaluation supplies the true descriptor, whereas estimated-CSI evaluation constructs it from the shared pilot observation described in Section~\ref{sec:setup}.

\paragraph{Descriptor exchange and buffering.} When the channel varies slowly relative to the payload transmission rate, the receiver can send its estimated descriptor $\widehat{\db}$ to the transmitter over a control link. The transmitter stores the latest shared descriptor in a condition buffer, and both endpoints reuse it for subsequent payload frames. After one or more frames, the receiver can refresh its estimate and send an updated descriptor; both endpoints switch to the new value at an agreed update boundary. In this way, condition feedback can be less frequent than payload transmission while the transmitter uses the buffered descriptor to generate each waveform.

\paragraph{Payload transmission.} For each payload, the fixed channel encoder produces $\bs{c}$, and the neural transmitter generates the complete complex waveform in one conditioned forward pass followed by frame-energy normalization. After propagation, the neural receiver directly stacks the real and imaginary parts of the received sequence, applies its conditioned forward pass, and outputs the coded-bit logits $\bs{\lambda}$. The fixed channel decoder then recovers the information bits. Thus, condition acquisition and descriptor exchange occur on the control path, whereas bit-to-waveform synthesis and bit recovery remain on the payload path.

\section{Comparison-model details}
\label{app:classical}

All methods use the same information and coded-bit lengths, channel distribution, transmit-energy normalization, and 490-sample frame budget. DeepOFDM uses its released two-dimensional residual-grid transmitter and receiver. DeepOFW is reconstructed from the released channel-conditioned waveform generator and uses five blocks of 80 waveform samples with an 18-sample cyclic prefix per block, ensuring that its guard interval covers the complete channel memory. The conventional OFDM baseline uses 32 subcarriers, 14 symbols, a three-sample cyclic prefix, QPSK, and LMMSE equalization; we report both perfect-CSI and pilot-estimated variants. Coded BER is measured directly from receiver logits, and information BER is measured after the same 20-iteration LDPC decoder for every method.

\raggedbottom
\section{Spectral response across CDL profiles and delay spreads}
\label{app:txchan-grid}

To examine whether the channel-dependent transmit-spectrum response in Figure~\ref{fig:wave-response} extends beyond the selected high-selectivity examples, we evaluate one random channel realization for each combination of CDL-A--E and RMS delay spreads of $0.10$, $0.25$, and $0.60\,\mu$s. We use the same universal power template $T_{\mathrm{raw}}(f)$ as in Section~\ref{sec:channel-adaptation}, estimated from 256 channels and 64 random-bit frames per channel. For each test channel at $10\,$dB $E_b/N_0$, we average the transmit power over 2048 random-bit frames on the native 490-bin whole-frame Fourier grid and compute $R_{\mathrm{dB}}(f)=10\log_{10}\!\left(P_{\hb}(f)/T_{\mathrm{raw}}(f)\right)$. We compare it with the channel power gain $G_{\mathrm{dB}}(f)=10\log_{10}|H(f)|^2$. After excluding the bins around DC and Nyquist, we mean-center both in-band vectors and calculate their Pearson correlation $r$, the regression slope of $R_{\mathrm{dB}}$ on $G_{\mathrm{dB}}$, and the gain span $\max G_{\mathrm{dB}}-\min G_{\mathrm{dB}}$. The metrics use unsmoothed data; only the displayed curves use 15-bin circular smoothing and fixed axes across panels.

\begin{figure}[H]
\centering
\includegraphics[width=0.95\textwidth,height=0.77\textheight,keepaspectratio]{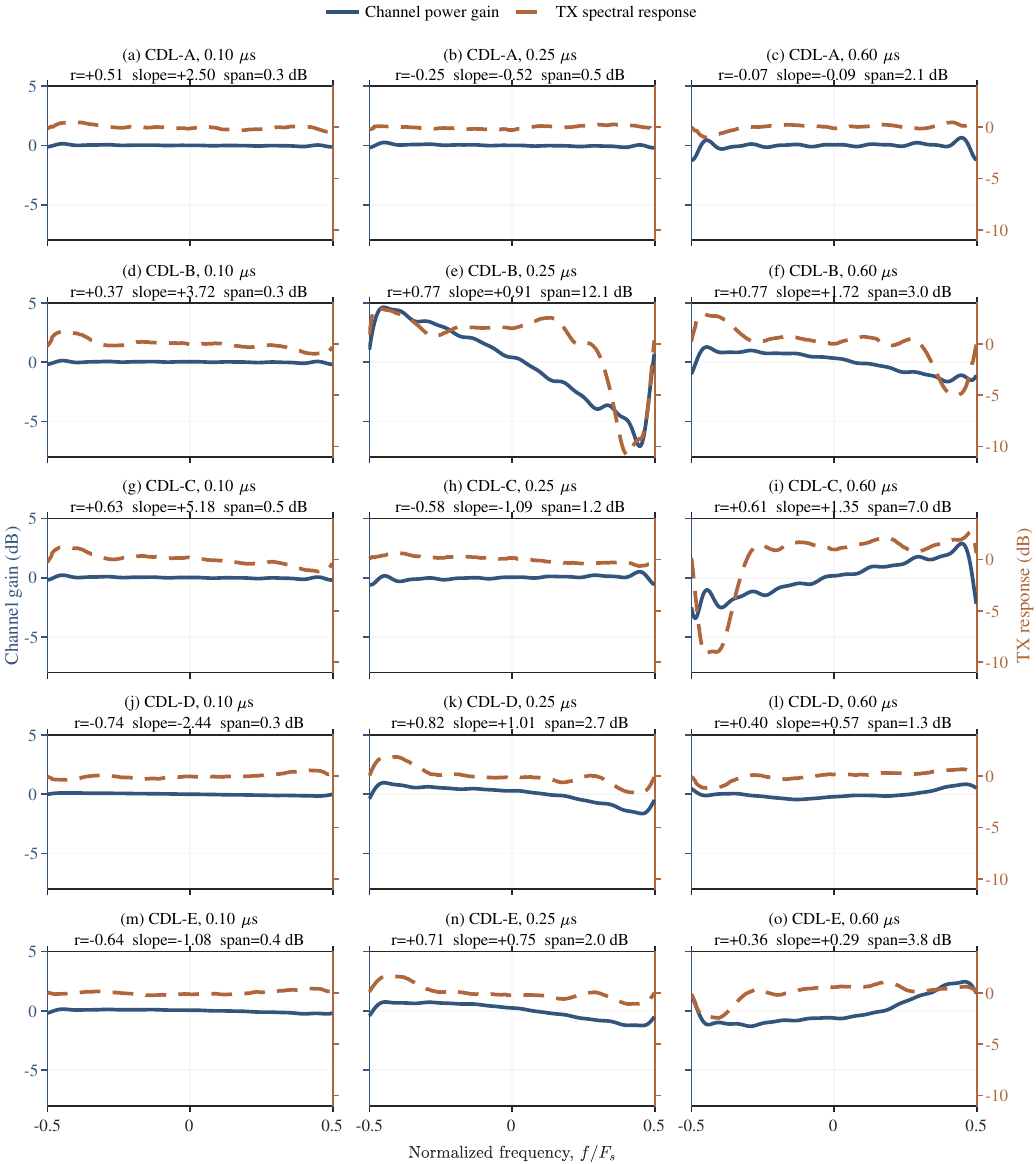}
\caption{Channel gain and template-normalized transmit-spectrum response across CDL-A--E (rows) and three RMS delay spreads (columns). Each panel shows one channel realization. Titles report the in-band Pearson correlation $r$, regression slope, and gain span computed before smoothing; the plotted curves are smoothed for readability and share fixed axis limits.}
\label{fig:appendix-txchan-grid}
\end{figure}

Figure~\ref{fig:appendix-txchan-grid} shows variation in the spectral response across individual channel realizations. Five of the seven panels with a gain span of at least $2\,$dB have $r\geq 0.6$, including CDL-B at $0.25\,\mu$s ($r=0.77$, span $12.1\,$dB) and CDL-D at $0.25\,\mu$s ($r=0.82$, span $2.7\,$dB). For CDL-A and CDL-E at $0.60\,\mu$s, the correlations are $-0.07$ and $0.36$, with gain spans of $2.1$ and $3.8\,$dB, respectively. Near-flat realizations with gain spans below $1\,$dB exhibit small spectral variations, making their correlations and slopes sensitive to these variations. The selected deep-fade realizations in Figure~\ref{fig:wave-response}(b)--(d) exhibit stronger gain--response correlations near $0.9$.

\section{Per-scale contribution of the multiscale synthesis heads}
\label{app:scale-contrib}

To examine how the trained model uses different temporal scales, we test the four branches of its synthesis head with their trained weights fixed. Normally, each branch converts one scale's tokens to a common latent length, and their outputs are combined before the final readout. We first activate one branch at a time and measure the resulting TX waveform energy or RX logit magnitude. The solo-branch shares normalize these isolated outputs across the four branches. We then remove one branch at a time and measure the change in BER to assess how much bit recovery depends on that branch. When all four branches are active, this test reproduces the original model output exactly.
\begin{figure}[H]
\centering
\includegraphics[width=0.96\textwidth,height=0.72\textheight,keepaspectratio]{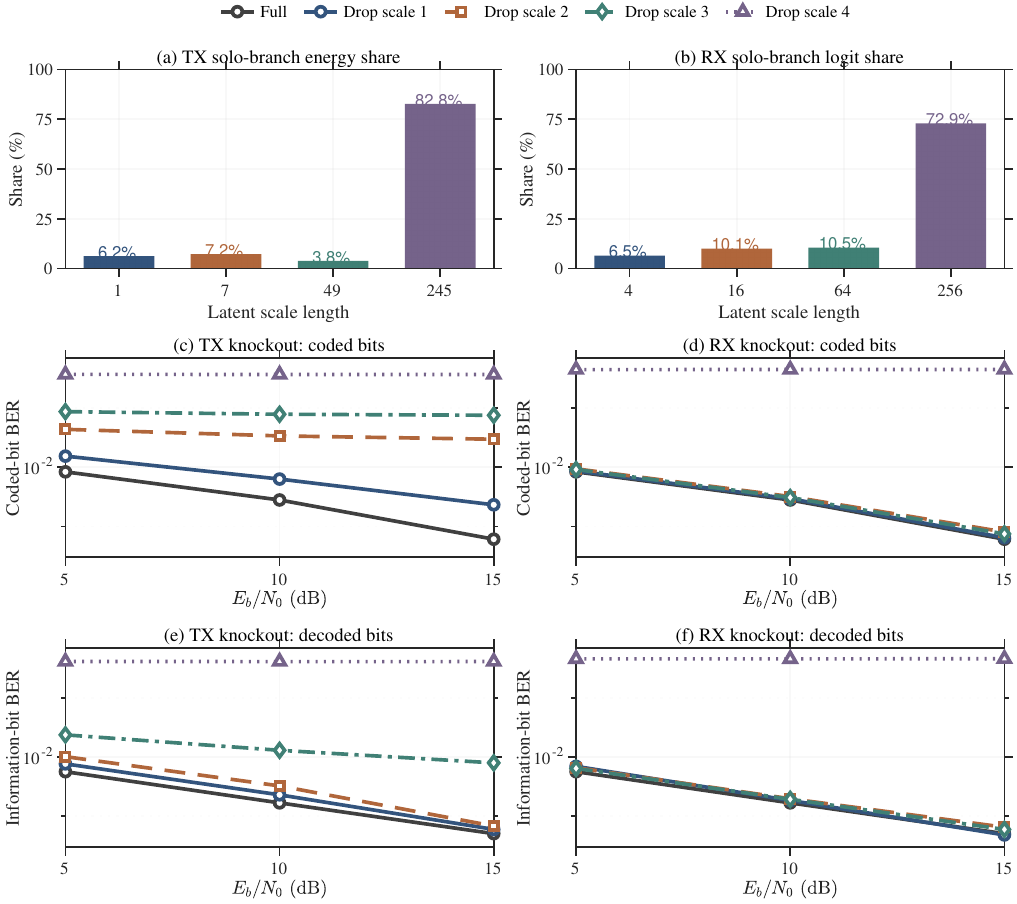}
\caption{Contribution of individual multiscale-head branches. (a,b) Normalized solo-branch TX waveform-energy and RX mean-absolute-logit shares. (c,d) Coded-bit BER and (e,f) post-LDPC information-bit BER when one TX or RX branch is removed. Scale 1--4 correspond to latent lengths $(1,7,49,245)$ at the TX and $(4,16,64,256)$ at the RX. All knockout curves use the same trained weights and paired channel/noise pool.}
\label{fig:appendix-scale-contrib}
\end{figure}

The knockout evaluation uses a shared pool of 3200 frames per $E_b/N_0$ point at $5$, $10$, and $15\,$dB, with CDL-A--E sampled uniformly and RMS delay spread sampled log-uniformly from $0.1$ to $0.6\,\mu$s. Figure~\ref{fig:appendix-scale-contrib}(a)--(b) reports the single-branch measurements: when operated separately, the finest TX and RX branches produce $82.8\%$ of the summed isolated TX waveform energy and $72.9\%$ of the summed isolated RX absolute-logit magnitude, respectively. Removing a branch tests a different property. At $15\,$dB, disabling the TX branches with latent lengths $1$, $7$, $49$, and $245$ increases coded-bit BER by factors of $3.8$, $49$, $125$, and $612$ relative to the full model. Thus, even coarse TX branches with small isolated outputs are important for bit recovery. The post-LDPC information-bit BER is less affected when either of the first two TX branches is removed, consistent with error correction mitigating part of the coded-bit degradation. On the RX side, disabling the first three branches increases coded-bit BER by only $6\%$, $32\%$, and $23\%$, whereas removing the finest branch raises it to approximately $0.45$. The trained RX readout is therefore much more sensitive to removal of the finest branch. These tests establish branch-level dependence.

\subsection{Waveform sensitivity to individual scales}
\label{app:scale-sensitivity}

We further probe the Transformer output immediately before the TX synthesis head. Let $\bs{t}_j=\operatorname{vec}(\bs{G}^{\mathrm{TX},(j)})$ denote the tokens at scale $j$, flattened into a vector, and let $\xb_{\mathrm{RI}}=[\operatorname{Re}(\xb)^{\mathsf T},\operatorname{Im}(\xb)^{\mathsf T}]^{\mathsf T}$ be the real representation of the transmitted waveform. We define the scale's gradient energy and its normalized share as
\begin{equation}
g_j=\Ebb\left\|\frac{\partial\xb_{\mathrm{RI}}}{\partial\bs{t}_j}\right\|_{\mathrm F}^{2},
\qquad
s_j=\frac{g_j}{\sum_{m=1}^{J}g_m},
\qquad
\bar{s}_j=\frac{g_j/\ell_j^{\mathrm{TX}}}{\sum_{m=1}^{J}g_m/\ell_m^{\mathrm{TX}}},
\label{eq:scale-gradient}
\end{equation}
where $\|\cdot\|_{\mathrm F}$ is the Frobenius norm and $\bar{s}_j$ normalizes for the number of tokens. We estimate these quantities from 512 frames at $10$\,dB, using static CDL-A--E channels with RMS delay spreads in $[0.1,0.6]\,\mu$s and perfect CSI.

To distinguish payload and channel sensitivity, we also replace one output token group with tokens produced by another codeword under the same channel, or by the same codeword under another channel. All other groups are retained. For the resulting waveform $\xb_{\mathrm{swap}}$, we measure the relative squared change $\Delta_x=\|\xb_{\mathrm{swap}}-\xb\|_2^2/\|\xb\|_2^2$, averaged over 32 channels and 32 codewords per channel. Table~\ref{tab:scale-sensitivity} reports both probes.

\begin{table}[!htbp]
\centering
\caption{TX scale sensitivity at $10$\,dB. Shares and relative squared waveform changes are shown as percentages.}
\label{tab:scale-sensitivity}
\small
\setlength{\tabcolsep}{6pt}
\begin{tabular}{@{}ccccc@{}}
\toprule
\shortstack{Latent\\length} & \shortstack{Gradient\\share $s_j$} & \shortstack{Per-token\\share $\bar{s}_j$} & \shortstack{$\Delta_x$: different\\codeword} & \shortstack{$\Delta_x$: different\\channel} \\
\midrule
1   & 24.4 & 89.1 & 35.1  & 1.0 \\
7   & 17.3 & 9.0  & 48.6  & 2.1 \\
49  & 17.2 & 1.3  & 53.5  & 3.8 \\
245 & 41.1 & 0.6  & 166.8 & 144.8 \\
\bottomrule
\end{tabular}
\end{table}

All four scales affect the synthesized waveform. The three coarser scales jointly account for $58.9\%$ of waveform-gradient energy, while the single coarsest token accounts for $24.4\%$. Replacing tokens with those of another payload changes the waveform at every scale. Replacing them across channels has a much larger effect at the finest scale ($144.8\%$) than at the coarser scales ($1.0$--$3.8\%$). These sensitivities, together with the branch-removal BER results, connect the learned multiscale representation to both physical waveform synthesis and bit recovery.

\section{Influence and perturbation analysis}
\label{app:influence}

\textbf{TX/RX influence.} To characterize the encoding and decoding dependencies learned by \method{}, we measure how much one output element changes when a single input element is perturbed, with the other inputs and environment descriptor held fixed. Let $B_{\mathrm{TX}}$ and $B_{\mathrm{RX}}$ denote the numbers of base coded-bit sequences used to probe the transmitter and received frames used to probe the receiver, respectively. At the TX, each coded bit is flipped individually. We denote the change in its centered input value by $\Delta_b$ and the resulting waveform samples for base sequence $i$, with and without flipping bit $j$, by $x_t^{(i,j)}$ and $x_t^{(i)}$. The influence of bit $j$ on complex transmit sample $t$, normalized by the input change, is
\begin{equation}
d_{\mathrm{TX}}(j,t)=\frac{1}{B_{\mathrm{TX}}\Delta_b}\sum_{i=1}^{B_{\mathrm{TX}}}\left|x_t^{(i,j)}-x_t^{(i)}\right|.
\end{equation}
At the RX, we add a complex Gaussian perturbation $\delta_t$ to one received sample at a time, with RMS amplitude $\rho\,\operatorname{RMS}(\yb^{(i)})$ for frame $i$. If $\lambda_j^{(i,t,\delta_t)}$ and $\lambda_j^{(i)}$ denote the coded-bit logits with and without this perturbation, the influence of received sample $t$ on logit $j$ is
\begin{equation}
d_{\mathrm{RX}}(j,t)=\frac{1}{B_{\mathrm{RX}}}\sum_{i=1}^{B_{\mathrm{RX}}}\mathbb{E}_{\delta_t}\left|\lambda_j^{(i,t,\delta_t)}-\lambda_j^{(i)}\right|,
\end{equation}
where the expectation is estimated from the sampled perturbations. In this experiment, $B_{\mathrm{TX}}=8$, $B_{\mathrm{RX}}=32$, and $\rho=0.05$. Since the centered TX input takes values $-\tfrac12$ and $+\tfrac12$, its bit-flip magnitude is $\Delta_b=1$. We use one fixed CDL-D channel (seed 700, $0.21\,\mu$s delay spread) at $10$\,dB $E_b/N_0$, with $C=768$ coded bits, $N=490$ transmit samples, and $N_{\mathrm{RX}}=508$ received samples. Flips and small additive perturbations probe the discrete TX input and continuous RX input, respectively; neither requires assumptions about the model internals.

\begin{figure}[!htbp]
\centering
\begin{subfigure}[t]{0.49\linewidth}
\centering
\includegraphics[width=\linewidth,height=1.70in,keepaspectratio]{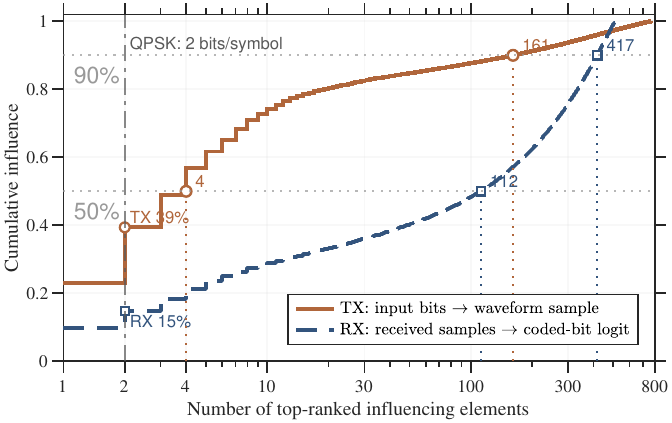}
\caption{Cumulative TX/RX influence.}
\label{fig:influence}
\end{subfigure}\hfill
\begin{subfigure}[t]{0.49\linewidth}
\centering
\includegraphics[width=\linewidth,height=1.70in,keepaspectratio]{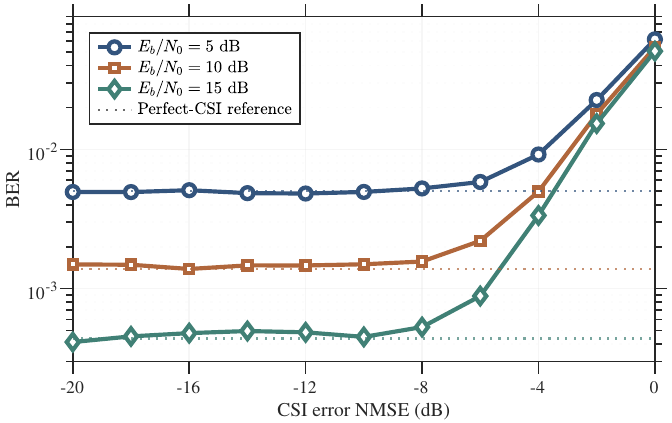}
\caption{CSI-condition perturbation.}
\label{fig:csi-perturbation}
\end{subfigure}
\caption{Complementary sensitivity analyses. (a) Coded bits are ranked by their effect on one transmitted sample, while received samples are ranked by their effect on one coded-bit logit; the rank-two marker provides a conceptual reference for the two bits carried by each QPSK constellation symbol. (b) BER under controlled perturbations of the injected CSI condition, with the physical channel and transmitted waveform held fixed.}
\label{fig:mechanism-sensitivity}
\end{figure}

For each of the $N$ transmit samples, we rank its $C$ bit influences in descending order, divide by their sum, and accumulate the top-$k$ shares; the TX curve is the median of these cumulative curves across samples. Likewise, for each of the $C$ logits, we rank and normalize the influences of the $N_{\mathrm{RX}}$ received samples, then take the median cumulative curve across logits. A common multiplicative normalization of either influence matrix cancels in these shares. Influence therefore measures output sensitivity to finite input perturbations. Figure~\ref{fig:influence} shows that four and 161 bits are needed to reach 50\% and 90\% of the median TX cumulative influence, respectively; the corresponding RX counts are 112 and 417 samples. These results indicate a distributed bit-to-waveform representation and broad aggregation across the received waveform.

\textbf{Sensitivity to CSI perturbations.} We further perturb the injected CSI while keeping the physical propagation channel fixed. As shown in Figure~\ref{fig:csi-perturbation}, G-Wave is largely insensitive to mild CSI errors and exhibits gradual performance degradation as the perturbation increases. The system remains operational even under severe CSI distortion approaching $0$\,dB NMSE, indicating strong tolerance to imperfect channel-condition information.

\section{Additional reliability metrics and baseline comparisons}
\label{app:additional-comparisons}

This appendix supplements the nine-point information-BER summary in Section~\ref{sec:main}. Unless specified otherwise, supplementary mean error rates average the $16$ integer $E_b/N_0$ points from $0$ to $15$\,dB with equal weights, using the mixed CDL-A--E distribution. The decoder, payload size, waveform length, sampling rate, and frame-energy constraint are unchanged. We report coded-bit BER before LDPC decoding, information-bit BER after decoding, and BLER, defined as the fraction of frames with at least one incorrectly recovered information bit.

\subsection{Detection performance before LDPC decoding}

Table~\ref{tab:coded-info-supplement} reports both BERs from the full 16-point evaluation records. \method{} has the lowest coded-bit and information-bit BER under both CSI settings. With perfect CSI, the coded-bit BER of DeepOFDM is $2.23$ times that of \method{}, while the information-bit ratio is $1.94$. Thus, improved detection is visible before channel decoding as well as after it.

\begin{table}[!htbp]
\centering
\caption{Mean BER on mixed CDL-A--E channels over the 16-point sweep.}
\label{tab:coded-info-supplement}
\small
\setlength{\tabcolsep}{5pt}
\begin{tabular}{@{}lcccc@{}}
\toprule
& \multicolumn{2}{c}{Perfect CSI} & \multicolumn{2}{c}{Estimated CSI} \\
\cmidrule(lr){2-3}\cmidrule(lr){4-5}
Method & \shortstack{Coded BER\\(before LDPC)} & \shortstack{Info BER\\(after LDPC)} & \shortstack{Coded BER\\(before LDPC)} & \shortstack{Info BER\\(after LDPC)} \\
\midrule
\method{} & $\boldsymbol{7.34\times10^{-3}}$ & $\boldsymbol{4.83\times10^{-3}}$ & $\boldsymbol{8.34\times10^{-3}}$ & $\boldsymbol{5.75\times10^{-3}}$ \\
DeepOFW & $9.08\times10^{-3}$ & $6.03\times10^{-3}$ & $1.30\times10^{-1}$ & $1.41\times10^{-1}$ \\
DeepOFDM & $1.64\times10^{-2}$ & $9.39\times10^{-3}$ & $1.62\times10^{-2}$ & $9.97\times10^{-3}$ \\
OFDM & $3.37\times10^{-2}$ & $1.03\times10^{-2}$ & $6.01\times10^{-2}$ & $2.11\times10^{-2}$ \\
\bottomrule
\end{tabular}
\end{table}

\subsection{Block errors and operating points}
\label{app:target-error-rates}

We evaluate 16,000 frames per SNR point for the supplementary BLER comparison. Target-error operating points are obtained by linear interpolation of log error rate between neighboring SNR points. A value of $>15$\,dB means that the target was not reached within the evaluated range. Table~\ref{tab:bler-supplement} reports mean BLER and the SNR needed for $1\%$ BLER. \method{} reaches this target $2.59$\,dB before DeepOFDM with perfect CSI and $2.90$\,dB before it with estimated CSI. The BCE-only DeepOFW and DeepOFDM variants are trained without a PAPR penalty; \emph{no conditioning} removes CSI conditioning from DeepOFDM.

\begin{table}[!htbp]
\centering
\caption{Mean BLER and the operating point at $1\%$ BLER. Gaps are relative to \method{} under the same CSI setting.}
\label{tab:bler-supplement}
\small
\setlength{\tabcolsep}{6pt}
\begin{tabular}{@{}lccc@{}}
\toprule
\multicolumn{4}{@{}l}{\textbf{Perfect CSI}} \\
\midrule
Method & Mean BLER & \shortstack{$E_b/N_0$ at\\$1\%$ BLER (dB)} & Gap (dB) \\
\midrule
\method{} & $\boldsymbol{1.90\times10^{-2}}$ & 7.83 & 0.00 \\
DeepOFW, BCE only & $2.24\times10^{-2}$ & 9.03 & 1.20 \\
DeepOFDM & $3.73\times10^{-2}$ & 10.42 & 2.59 \\
DeepOFDM, no conditioning & $5.25\times10^{-2}$ & 12.24 & 4.41 \\
DeepOFDM, BCE only & $3.12\times10^{-2}$ & 10.06 & 2.23 \\
OFDM-LMMSE & $4.37\times10^{-2}$ & 12.31 & 4.48 \\
\bottomrule
\end{tabular}

\medskip
\begin{tabular}{@{}lccc@{}}
\toprule
\multicolumn{4}{@{}l}{\textbf{Estimated CSI}} \\
\midrule
Method & Mean BLER & \shortstack{$E_b/N_0$ at\\$1\%$ BLER (dB)} & Gap (dB) \\
\midrule
\method{} & $\boldsymbol{2.10\times10^{-2}}$ & 8.25 & 0.00 \\
DeepOFW & $4.41\times10^{-1}$ & $>15$ & \textemdash \\
DeepOFDM & $3.89\times10^{-2}$ & 11.15 & 2.90 \\
OFDM-LS & $9.53\times10^{-2}$ & $>15$ & \textemdash \\
\bottomrule
\end{tabular}
\end{table}

Table~\ref{tab:ber-target-supplement} additionally reports the SNR needed for an information-bit BER of $10^{-3}$. The corresponding advantage over DeepOFDM is $2.14$\,dB with perfect CSI and approximately $1.50$\,dB with estimated CSI.

\begin{table}[!htbp]
\centering
\caption{$E_b/N_0$ (dB) required for information-bit BER $10^{-3}$.}
\label{tab:ber-target-supplement}
\small
\begin{tabular}{@{}lcc@{}}
\toprule
Method & Perfect CSI & Estimated CSI \\
\midrule
\method{} & 12.26 & 12.44 \\
DeepOFDM & 14.40 & 13.93 \\
DeepOFW & 12.85 & $>15$ \\
OFDM & 14.95 & $>15$ \\
\bottomrule
\end{tabular}
\end{table}

\subsection{CSI conditioning and water-filling}
\label{app:water-filling}

We evaluate OFDM with CSI-based water-filling~\citep{tse2005fundamentals} to compare direct waveform synthesis with explicit frequency-domain power adaptation. This baseline uses the same frame and energy constraints, with a fixed payload, QPSK modulation, and LDPC code. Table~\ref{tab:water-filling} shows that \method{} achieves lower mean information-bit BER than water-filling OFDM under both perfect and estimated CSI, demonstrating a performance advantage over this explicit power-allocation baseline.

\begin{table}[!htbp]
\centering
\caption{Mean information-bit BER for OFDM with and without CSI-based water-filling over the 16-point sweep.}
\label{tab:water-filling}
\small
\setlength{\tabcolsep}{12pt}
\begin{tabular}{@{}lcc@{}}
\toprule
Method & Perfect CSI & Estimated CSI \\
\midrule
\method{} & $\boldsymbol{4.83\times10^{-3}}$ & $\boldsymbol{5.75\times10^{-3}}$ \\
OFDM & $1.03\times10^{-2}$ & $2.11\times10^{-2}$ \\
OFDM + water-filling & $1.07\times10^{-2}$ & $2.44\times10^{-2}$ \\
\bottomrule
\end{tabular}
\end{table}

We also train DeepOFDM without CSI conditioning. Its mean information-bit BER is $1.33\times10^{-2}$, compared with $9.22\times10^{-3}$ for the no-conditioning \method{} ablation in Table~\ref{tab:ablation}. Conditioning is beneficial, and its implementation also matters: replacing AdaLN-Zero by concatenative conditioning increases the \method{} ablation BER from $4.57\times10^{-3}$ to $5.24\times10^{-3}$. The frequency-domain and resource-grid variants in that table separately evaluate the effect of constraining the learned waveform interface.

\subsection{Diffusion-assisted channel estimation}
\label{app:diffusion-estimation}

We integrate the generative diffusion estimator of \citet{zhou2025diffusion} into the OFDM receiver. Following the authors' released code, we re-implement the denoising network, noise schedule, training recipe, and posterior inference in TensorFlow. We adapt the original two-dimensional architecture to a one-dimensional delay-domain channel vector for the single-antenna, 32-subcarrier task and retrain it on our channel distribution. The estimator uses 100 diffusion timesteps with a linear noise-variance schedule from $10^{-4}$ to $0.1$, noise prediction, and Adam with learning rate $10^{-4}$ and batch size 128. Inference hyperparameters are selected using validation NMSE.

This additional comparison uses a common evaluation pool with 16,000 frames per SNR point over $0$--$15$\,dB. Table~\ref{tab:diffusion-estimation} reports results from this pool, independently of the original aggregate records in Table~\ref{tab:coded-info-supplement}. Normalized mean-square error (NMSE) measures channel-estimation error relative to channel energy; a lower value is better. \method{} and DeepOFDM receive the same LS-based estimate, so their descriptor NMSE is that of LS estimation. The diffusion estimator improves NMSE from $-7.6$ to $-12.9$\,dB and reduces OFDM information-bit BER from $2.27\times10^{-2}$ to $1.44\times10^{-2}$. In the same comparison, \method{} achieves $5.46\times10^{-3}$, demonstrating the benefit of the learned transceiver relative to this diffusion-assisted OFDM pipeline.

\begin{table}[!htbp]
\centering
\caption{Diffusion-estimation comparison on its shared evaluation pool. Coded BER is measured before LDPC decoding.}
\label{tab:diffusion-estimation}
\scriptsize
\setlength{\tabcolsep}{4.5pt}
\begin{tabular}{@{}lccccc@{}}
\toprule
Method & \shortstack{NMSE\\(dB)} & Coded BER & Info BER & BLER & \shortstack{$E_b/N_0$ at\\$1\%$ BLER (dB)} \\
\midrule
\method{}, LS CSI & $-7.6$ & $\boldsymbol{8.05\times10^{-3}}$ & $\boldsymbol{5.46\times10^{-3}}$ & $\boldsymbol{2.10\times10^{-2}}$ & 8.25 \\
DeepOFDM, LS CSI & $-7.6$ & $1.64\times10^{-2}$ & $1.00\times10^{-2}$ & $3.89\times10^{-2}$ & 11.15 \\
OFDM + diffusion & $-12.9$ & $4.01\times10^{-2}$ & $1.44\times10^{-2}$ & $7.33\times10^{-2}$ & $>15$ \\
OFDM + LS & $-7.6$ & $6.13\times10^{-2}$ & $2.27\times10^{-2}$ & $9.95\times10^{-2}$ & $>15$ \\
OFDM, perfect CSI & $-\infty$ & $3.45\times10^{-2}$ & $1.12\times10^{-2}$ & $4.65\times10^{-2}$ & 12.23 \\
\bottomrule
\end{tabular}
\end{table}

\subsection{Peak power and baseline training objectives}
\label{app:papr}

We additionally evaluate peak-to-average power ratio (PAPR) using fourfold oversampling. For each oversampled frame, PAPR is the maximum sample power divided by its mean sample power, expressed in dB. We report its mean and thresholds exceeded by $1\%$ and $0.1\%$ of frames, corresponding to complementary cumulative distribution function (CCDF) levels $10^{-2}$ and $10^{-3}$. Frames from the $0$--$15$\,dB evaluation are pooled.

The reconstructed DeepOFW comparison is trained using BCE without a PAPR penalty. For DeepOFDM, we evaluate both the original objective with PAPR regularization and a BCE-only variant. Table~\ref{tab:papr-supplement} shows that \method{} has lower mean and tail PAPR than the evaluated baselines. Its mean is $7.32$\,dB, versus $8.73$\,dB for BCE-only DeepOFW; the $10^{-3}$ CCDF thresholds are $10.76$ and $11.52$\,dB. The error-rate comparison in Table~\ref{tab:bler-supplement} also includes these objective controls: mean perfect-CSI BLER is $1.90\times10^{-2}$ for \method{}, $2.24\times10^{-2}$ for BCE-only DeepOFW, and $3.12\times10^{-2}$ for BCE-only DeepOFDM.

\begin{table}[!htbp]
\centering
\caption{PAPR with fourfold oversampling (dB).}
\label{tab:papr-supplement}
\small
\setlength{\tabcolsep}{9pt}
\begin{tabular}{@{}lccc@{}}
\toprule
Method & Mean & CCDF $10^{-2}$ & CCDF $10^{-3}$ \\
\midrule
\method{} & 7.32 & 10.16 & 10.76 \\
DeepOFW, BCE only & 8.73 & 10.64 & 11.52 \\
DeepOFDM, with PAPR penalty & 10.34 & 11.59 & 11.82 \\
DeepOFDM, BCE only & 10.58 & 12.56 & 13.34 \\
OFDM (QPSK) & 8.51 & 10.33 & 10.99 \\
\bottomrule
\end{tabular}
\end{table}

\FloatBarrier
\section{Generalization to unseen sites and mobile channels}
\label{app:generalization}

\subsection{Site-specific ray-tracing channels}

We evaluate two site-specific propagation datasets, Sionna RT Munich~\citep{sionnaRTintro} and DeepMIMO ASU campus~\citep{deepmimoQuickstart}. Their ray-traced propagation paths provide channel realizations tied to scene geometry. All learned methods use their original checkpoints, without training or fine-tuning on either dataset. We use perfect CSI and 16,000 frames at each integer $E_b/N_0$ from $0$ to $15$\,dB, retaining the link interfaces and decoder of Section~\ref{sec:setup}.

Table~\ref{tab:ray-tracing} shows that \method{} achieves the lowest mean coded-bit and information-bit BER on both datasets. Relative to the strongest information-BER baseline on each site, the reductions are approximately $46.9\%$ on Munich (DeepOFDM) and $49.0\%$ on ASU (DeepOFW). These results extend the evaluation from statistical CDL profiles to unseen site-specific ray-tracing environments.

\begin{table}[!htbp]
\centering
\caption{Mean BER on unseen ray-tracing channels with perfect CSI.}
\label{tab:ray-tracing}
\small
\setlength{\tabcolsep}{8pt}
\begin{tabular}{@{}lcccc@{}}
\toprule
& \multicolumn{2}{c}{Sionna RT Munich} & \multicolumn{2}{c}{DeepMIMO ASU campus} \\
\cmidrule(lr){2-3}\cmidrule(lr){4-5}
Method & Coded BER & Info BER & Coded BER & Info BER \\
\midrule
\method{} & $\boldsymbol{2.19\times10^{-3}}$ & $\boldsymbol{1.20\times10^{-3}}$ & $\boldsymbol{4.94\times10^{-3}}$ & $\boldsymbol{2.77\times10^{-3}}$ \\
DeepOFW & $4.57\times10^{-3}$ & $3.30\times10^{-3}$ & $8.41\times10^{-3}$ & $5.43\times10^{-3}$ \\
DeepOFDM & $7.72\times10^{-3}$ & $2.26\times10^{-3}$ & $1.43\times10^{-2}$ & $5.93\times10^{-3}$ \\
OFDM & $2.45\times10^{-2}$ & $2.63\times10^{-3}$ & $3.14\times10^{-2}$ & $6.85\times10^{-3}$ \\
\bottomrule
\end{tabular}
\end{table}

\subsection{Mobility with estimated CSI}
\label{app:mobility}

We retain the static-trained checkpoints and evaluate time-varying CDL channels at user speeds from $0$ to $108$\,km/h. The carrier frequency remains $3.5$\,GHz, and each speed--SNR pair uses 16,000 frames. Both methods receive estimated CSI under the same observation protocol as Section~\ref{sec:setup}. For this experiment, each reported mean averages the information-bit BER at $E_b/N_0\in\{5,10,15\}$\,dB.

As shown in Table~\ref{tab:mobility}, \method{} has lower mean BER than DeepOFDM at every tested speed through $72$\,km/h. At $72$\,km/h, the means are $3.48\times10^{-3}$ and $4.94\times10^{-3}$, respectively. At $108$\,km/h, the corresponding means are $4.06\times10^{-2}$ and $1.78\times10^{-2}$. These results demonstrate transfer from static training to the evaluated mobility conditions through $72$\,km/h.

\begin{table}[!htbp]
\centering
\caption{Mean information-bit BER under mobility with estimated CSI, averaged over $5$, $10$, and $15$\,dB.}
\label{tab:mobility}
\small
\setlength{\tabcolsep}{16pt}
\begin{tabular}{@{}ccc@{}}
\toprule
Speed (km/h) & \method{} & DeepOFDM \\
\midrule
0   & $\boldsymbol{2.72\times10^{-3}}$ & $4.72\times10^{-3}$ \\
9   & $\boldsymbol{2.85\times10^{-3}}$ & $4.97\times10^{-3}$ \\
18  & $\boldsymbol{2.67\times10^{-3}}$ & $4.80\times10^{-3}$ \\
27  & $\boldsymbol{2.65\times10^{-3}}$ & $4.39\times10^{-3}$ \\
36  & $\boldsymbol{2.84\times10^{-3}}$ & $5.01\times10^{-3}$ \\
54  & $\boldsymbol{2.95\times10^{-3}}$ & $4.58\times10^{-3}$ \\
72  & $\boldsymbol{3.48\times10^{-3}}$ & $4.94\times10^{-3}$ \\
108 & $4.06\times10^{-2}$ & $\boldsymbol{1.78\times10^{-2}}$ \\
\bottomrule
\end{tabular}
\end{table}

\FloatBarrier
\section{Computational complexity and memory}
\label{app:complexity}

We profile TX and RX inference together for a 490-sample frame carrying 384 information bits, excluding channel simulation and LDPC decoding. Table~\ref{tab:complexity-supplement} reports parameter counts, floating-point operations, and single-frame latency on an NVIDIA A100. XLA compilation is used for \method{}, DeepOFDM, and OFDM-LMMSE. DeepOFW is profiled using its optimized, non-XLA GRU implementation. Its shared parameter-generating network is counted once.

\method{} uses $34.0\%$ fewer parameters and $46.0\%$ of the operations of DeepOFDM. Its compiled latency is $3.68$\,ms, compared with $3.23$\,ms for DeepOFDM. The DeepOFW and conventional OFDM implementations take $2.00$ and $0.69$\,ms, respectively. Together with the reliability results, these measurements characterize the performance--cost trade-off across transceiver architectures.

\begin{table}[!htbp]
\centering
\caption{Model complexity and A100 inference latency at batch size one.}
\label{tab:complexity-supplement}
\small
\setlength{\tabcolsep}{7pt}
\begin{tabular}{@{}lcccl@{}}
\toprule
Method & Parameters (M) & GFLOP/frame & Latency (ms) & Execution \\
\midrule
\method{} & 17.59 & 11.0 & 3.68 & XLA \\
DeepOFW & 17.68 & 0.07 & 2.00 & Optimized GRU \\
DeepOFDM & 26.67 & 23.9 & 3.23 & XLA \\
OFDM-LMMSE & 0 & $6\times10^{-5}$ & 0.69 & XLA \\
\bottomrule
\end{tabular}
\end{table}

Table~\ref{tab:memory-supplement} reports FP32 weight storage and additional GPU memory allocated during non-XLA inference at batch sizes one and 64. \method{} uses $67$\,MB for weights, compared with $102$\,MB for DeepOFDM. Its additional inference allocations are $9$ and $594$\,MB at the two batch sizes; the corresponding DeepOFDM values are $10$ and $242$\,MB.

\begin{table}[!htbp]
\centering
\caption{FP32 weight storage and additional inference GPU memory (non-XLA; reported in MB).}
\label{tab:memory-supplement}
\small
\setlength{\tabcolsep}{10pt}
\begin{tabular}{@{}lccc@{}}
\toprule
Method & \shortstack{FP32\\weights} & \shortstack{Extra GPU memory\\batch size 1} & \shortstack{Extra GPU memory\\batch size 64} \\
\midrule
\method{} & 67 & 9 & 594 \\
DeepOFW & 68 & 47 & 122 \\
DeepOFDM & 102 & 10 & 242 \\
OFDM-LMMSE & 0 & 0 & 2 \\
\bottomrule
\end{tabular}
\end{table}

\FloatBarrier
\section{Learning objective and conditional waveform synthesis}
\label{app:learning-objective}

\paragraph{From coded bits to a waveform distribution.}
For a fixed descriptor $\db$, the transmitter maps random coded-bit sequences through bit projection, condition-modulated latent processing, and waveform synthesis. Let $g_\theta(\bs{c},\db)$ denote the complete TX mapping, including the power normalization in Equation~\ref{eq:link-tx}, and let $P_{\bs{c}}$ be the distribution induced by encoding the random information bits. For any measurable set $\mathcal A$ of complex waveforms, the induced conditional distribution is
\begin{equation}
P_\theta(\mathcal A\mid\db)
=
\Pr_{\bs{c}\sim P_{\bs{c}}}\!\left[g_\theta(\bs{c},\db)\in\mathcal A\right].
\label{eq:conditional-waveform-distribution}
\end{equation}
The learned condition-dependent map determines how payload information is realized in waveform space. Its supervision comes from successful bit recovery through the channel, so no target waveform distribution or waveform labels are required.

\paragraph{BCE and receiver uncertainty.}
We use the notation of Section~\ref{sec:preliminaries}: $\bs{b}$ contains $K$ independent uniform information bits, and the fixed injective encoder produces $C$ coded bits $\bs{c}=f_{\mathrm{enc}}(\bs{b})$. Coding introduces dependence among the $C$ bits, while preserving the information entropy $H(\bs{c})=H(\bs{b})=K\ln 2$. The descriptor is independent of the payload. All entropies and logarithms in this analysis are in natural units.

Let $q_{\phi,j}(1\mid\yb,\db)=\sigma(\lambda_j)$ be the receiver's probability for coded bit $j$, where $\sigma(u)=1/(1+\exp(-u))$, and let $p_{\theta,j}$ be the corresponding true conditional bit posterior induced by the transmitter and channel. The population BCE per coded bit satisfies
\begin{align}
C\mathcal L_{\mathrm{BCE}}
&=\sum_{j=1}^{C}\Ebb[-\ln q_{\phi,j}(c_j\mid\yb,\db)] \nonumber\\
&=\sum_{j=1}^{C}H(c_j\mid\yb,\db)
+\sum_{j=1}^{C}\Ebb_{\yb,\db}
D_{\mathrm{KL}}\!\left(p_{\theta,j}(\cdot\mid\yb,\db)\,\|\,q_{\phi,j}(\cdot\mid\yb,\db)\right),
\label{eq:bce-decomposition}
\end{align}
where $D_{\mathrm{KL}}$ is Kullback--Leibler divergence. Thus the objective combines residual bit uncertainty with receiver posterior mismatch. Conditional-entropy subadditivity and nonnegativity of KL divergence give
\begin{equation}
H(\bs{c}\mid\yb,\db)\leq
\sum_{j=1}^{C}H(c_j\mid\yb,\db)
\leq C\mathcal L_{\mathrm{BCE}}.
\label{eq:bce-entropy-bound}
\end{equation}
Because the code is injective, the information bits and their codeword contain the same information about the observation. Hence
\begin{equation}
I(\bs{b};\yb\mid\db)
=
I(\bs{c};\yb\mid\db)
\geq \max\!\left\{0,\ K\ln2-C\mathcal L_{\mathrm{BCE}}\right\}.
\label{eq:bce-mi-bound}
\end{equation}
Minimizing population BCE raises this lower bound on recovered information at the fixed payload size. The two inequalities in Equation~\ref{eq:bce-entropy-bound} are tight when the coded bits are conditionally independent given the received signal and descriptor and when the receiver predicts the true bit posteriors, respectively.

\paragraph{Relation to the channel constraint.}
Let $T_{\mathrm{fr}}$ be the frame duration and let $C_{\mathrm{Sh}}$ denote an upper bound, in bits per second, on achievable information transfer under the same channel, power, bandwidth, and available side information~\citep{tse2005fundamentals,goldsmith1997capacity}. Data processing gives
$I(\bs{b};\yb\mid\db)\leq T_{\mathrm{fr}}C_{\mathrm{Sh}}\ln2$.
Combining this with Equation~\ref{eq:bce-mi-bound} yields
\begin{equation}
\mathcal L_{\mathrm{BCE}}
\geq
\frac{\ln2}{C}
\left[K-T_{\mathrm{fr}}C_{\mathrm{Sh}}\right]_+,
\qquad [a]_+=\max\{a,0\}.
\label{eq:bce-capacity-bound}
\end{equation}
Here $K=384$ and $C=768$; the objective is normalized by coded bits, whereas the payload entropy is determined by information bits. The bound relates the training objective to recoverable information under fixed resources. Channel conditioning changes the waveform map and the resulting conditional uncertainty. The BCE decomposition connects waveform synthesis and bit recovery through two quantities optimized jointly: conditional bit uncertainty and receiver posterior mismatch.

\end{document}